\documentclass[11pt,a4paper]{article} 

\usepackage{adjustbox}
\usepackage{amsfonts}
\usepackage{amssymb}
\usepackage{amsmath}
\usepackage{amsthm}
\usepackage{authblk}
\usepackage{array}
\usepackage{bbm}
\usepackage{booktabs}
\usepackage{caption}
\usepackage{subcaption}
\usepackage{dsfont}
\usepackage{enumerate}
\usepackage{framed,multirow}
\usepackage{graphics}
\usepackage{graphicx}
\usepackage{hyperref}
\usepackage{inputenc}
\usepackage{latexsym}
\usepackage{makecell}
\usepackage{mathrsfs}  
\usepackage[table]{xcolor}
\usepackage{tikz}
\usepackage[normalem]{ulem}

\def\OOO{\mathcal{O}}

\theoremstyle{definition}

\title{Fit4CAD: A point cloud benchmark for fitting simple geometric primitives in CAD objects}

\author[1,\thanks{Corresponding author}]{Chiara Romanengo}
\author[1,2]{Andrea Raffo}
\author[3]{Yifan Qie}
\author[3]{Nabil Anwer}
\author[1]{Bianca Falcidieno}

\affil[1]{Istituto di Matematica Applicata e Tecnologie Informatiche  ``E. Magenes", Consiglio Nazionale delle Ricerche, Via de Marini 6, 16149 Genova, Italy.}
\affil[2]{Department of Mathematics, University of Oslo, Moltke Moes vei 35, 0851 Oslo, Norway.}
\affil[3]{Automated Production Research Laboratory (LURPA), ENS Paris-Saclay, Université Paris-Saclay, 91190 Gif-sur-Yvette, France.}

\date{}                     
\begin{document}
\maketitle

\begin{abstract}
We propose Fit4CAD, a benchmark for the evaluation and comparison of methods for fitting simple geometric primitives in point clouds representing CAD objects. This benchmark is meant to help both method developers and those who want to identify the best performing tools.
The Fit4CAD dataset is composed by 225 high quality point clouds, each of which has been obtained by sampling a CAD object. The way these elements were created by using existing platforms and datasets makes the benchmark easily expandable. The dataset is already split into a training set and a test set.
To assess performance and accuracy of the different primitive fitting methods, various measures are defined.
To demonstrate the effective use of Fit4CAD, we have tested it on two methods belonging to two different categories of approaches to the primitive fitting problem: a clustering method based on a primitive growing framework and a parametric method based on the Hough transform.
\end{abstract}

\section{Introduction}
3D CAD models are among the most common medium to convey dimensional and geometric information on designed objects or components. However, often the CAD model of an object is not available, it does not even exist, or no longer corresponds to the real geometry of the manufactured object itself. One strategy for retrieving a digital model of an object when not accessible is to acquire 3D data directly on the object and use it to create a digital representation. The reconstruction of digital models starting from the measured data is a process, commonly called Reverse Engineering (RE), aiming at  reconstructing 3D mathematical surfaces and geometric features that represent the geometry of real parts. Many methods have been proposed  to solve this problem; as a reference we cite a recent survey that groups a large part of the approaches presented so far \cite{ Kaiser2019}.

Given the large number of methods proposed, it becomes important to be able to evaluate their performance by creating standard datasets with a ground truth and a ``quality label", thus paving the road for a fair evaluation of the existing technologies and the identification of open research directions not only in reverse engineering but also in shape retrieval, understanding, compression, etc., taking inspiration from other approaches  proposed for generic classes of objects,  (e.g. \cite{Tal2007,PointNet,3DV2018}).

Here we propose Fit4CAD, a benchmark of point clouds representing CAD objects aimed at evaluating methods for detecting simple (polynomial) geometric primitives (i.e., plane, cylinder, cone, sphere, and torus) in 3D point clouds; by polynomial primitive, we here mean a surface that has an algebraic implicit representation, i.e., it can be defined as the zero set of a polynomial. The dataset consists of 225  high  quality point clouds, each of which has been obtained by sampling a CAD object. 
Each point cloud is equipped of a ground-truth segmentation and, for each primitive, we provide both implicit and parametric forms.
The way these elements were created by using existing platforms and datasets makes the benchmark easily expandable. Fit4CAD is designed to be used also by machine learning methods: in fact, the dataset comes in the form of a training set and a test set.

We provide a number of performance measures able to evaluate both the quality of the fitting segments, in term of points correctly recognized as belonging to a primitive, and the quality of the primitive approximation, evaluating the distance between the primitive detected and the ideal one.

Fit4CAD satisfies a certain number of necessary requirements, such as the relevance and representativeness of the elements in the CAD context, the richness and the completeness of the information associated with each primitive, thus enabling fair comparisons for a wide range of geometric primitives recognition algorithms.

The proposed benchmark has been exploited to evaluate and compare  two  methods of geometric primitive fitting,  belonging to two different categories of approaches: a clustering method based on a primitive growing framework and a parametric method based on the Hough transform. By providing an explicit representation of the equations of the primitives, for the second method we are also able to evaluate measures related to the accuracy of the primitives found.

The rest of the paper is organized as follows. Section \ref{sec:star} examines previous work related to our topic.
Section \ref{sec:benchmark} describes the characteristics of the benchmark: dataset, ground truth, and the performance and accuracy measures chosen to evaluate the identification of primitives.
Section \ref{sec:methods} describes the tests carried out on two methods of recognition and fitting of geometric primitives. Some concluding remarks end the paper.

\section{Prior work}
\label{sec:star}
Benchmarking involves sharing of resources,  metrics, data and so on, so that the common goals of knowledge creation and furthering the state of the art can be achieved. The creation of standard datasets reduces the amount of work necessary for single researchers to assess the quality of their techniques and compare them with other research groups. The steadily rising participation to contests and open challenges shows the interest and the need for benchmarks (e.g., TreCVID \cite{TrecVID}), competitive contests (e.g., on Kaggle.com \cite{Bojer_2021} or the 3D Shape Retrieval Contest (SHREC) \cite{SHREC2006}) and, more in general, for code sharing (e.g., Graphics Replicability Stamp Initiative\footnote{http://www.replicabilitystamp.org}).
So far, benchmarks for 3D object segmentation \cite{Chen2009,3DOR2012} have mainly considered generic classes of objects and, therefore, the methods were evaluated for their general-purpose segmentation rather than on CAD objects and their capability of recognizing geometric primivites. 

Among the datasets containing general 3D shapes (e.g., toys, mechanisms, jewelry) in the form of triangle meshes it is worth mentioning \cite{Thingi10k,Wu:2015,VISIONAIR}, even if these methods were specifically designed for different goals, e.g. 3D printing, computer vision applications and computer graphics applications, respectively.

The most relevant dataset for our work is the ABC dataset \cite{Koch_2019}; here, the authors present a massive dataset (over one million models),  specifically developed to train data-driven algorithms for geometric deep learning. Models are defined by parametric surfaces, possibly accompanied with the information  related to the decomposition into patches, sharp feature annotations, and analytic differential properties.
The models were created by using the interface available on the online  infrastructure Onshape\footnote{https://www.onshape.com/en/}. All models are stored as triangle meshes, while the associated files (annotations, features, etc.) are not available for all models: more precisely, the file may lack the true list of primitives, or their parametric/implicit representations; this is not a big issue for that specific dataset, as primitive extraction is not their purpose. Being specifically designed for data-driven methods, the models are stored with different resolutions (i.e., different samplings of the same parametric model) or slight variations. Moreover, models in \cite{Koch_2019} do not present any kind of data perturbation. Lastly, the ABC dataset is not designed for point cloud segmentation, and does not present any specific quality measures for comparing methods.

\section{The benchmark}
\label{sec:benchmark}
We here introduce our benchmark, geared towards the following desirable properties:
\begin{itemize}

    \item \emph{Dataset richness and representativeness}. The first and foremost requirement for a thorough evaluation is the availability of a data set characterized by a good sampling of varied shapes: each family of simple geometric primitives (i.e., plane, cylinder, cone, sphere, and torus) should appear in a sufficient number of point clouds. In addition, we consider different point cloud densities as well as data integrity (with/without missing data). Each CAD object is used to generate one and only one point cloud.
    
    \item \emph{Ease of expansion}. An ideal benchmark should be able to develop over time, in order to test new paradigms and face new challenges; this requires the capability to generate new data in an easy and efficient way. To satisfy this basic requirement, Section \ref{sec:dataset} outlines a general pipeline that can be used for data generation.
    
    \item \emph{Availability of both implicit and parametric representations}. Modern CAD systems are based on two complementary representations for surfaces, according to the manipulation they are involved in: implicit and parametric representations. Parametrized surfaces are best suited for point generation, while implicit representations allow to check whether a query point lies or not on the surface in a more convenient way. Having both representations makes it possible to answer a wide range of questions (e.g., intersection problems).
    
    \item \emph{Completeness of the documentation}. All models are equipped with all and the same information. For each of them, this includes:  the files with primitive segments, implicit and parametric primitive representations; a preset split of the dataset into training set and test set. Further details are provided in Section \ref{sec:ground}.
    
    \item \emph{Variety of performance indicators and accuracy measures}. To evaluate and compare methods, it is of vital importance to select measures that highlight strengths and weaknesses. In our case, the problem is twofold: on the one hand, we want to quantify the capability to produce precise segmentations into simple geometric primitives (by performance indicators); on the other hand, we also aim at measuring the fitting accuracy when it comes to implicit and parametric representation of the same shapes (by accuracy measures). 
    Section \ref{sec:metrics} describes the measures  selected to evaluate the detection of simple primitives in CAD objects point clouds.
\end{itemize}
 
 \subsection{Dataset}
 \label{sec:dataset}
\begin{figure}[h!]
     \begin{center}
     \begin{tabular}{cc}
          \includegraphics[scale=0.425, trim={0.5cm 0.10cm 0.5cm 0.5cm}, clip]{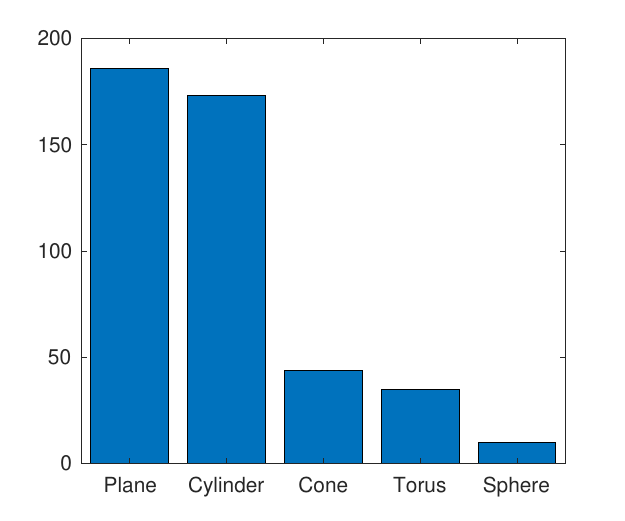}
          &
          \includegraphics[scale=0.425, trim={0.5cm 0.10cm 0.5cm 0.5cm}, clip]{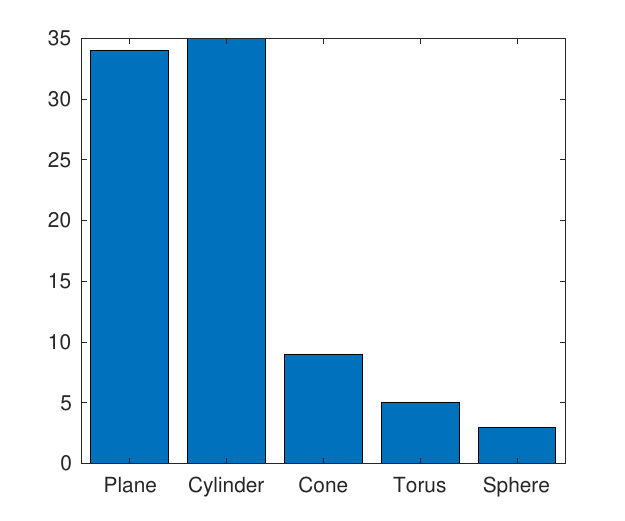}
     \end{tabular}
     \end{center}
     \caption{Surface type distribution. The two bar charts show the distributions for the training set (left) and the test set (right).}
     \label{fig:barchart_distributions}
 \end{figure}

 At present, the dataset contains $225$ individual high quality point clouds, each of which has been obtained by sampling a CAD object.  The dataset is already split into two subsets: a training set, counting $190$ point clouds, and a test set, containing the remaining $35$ point clouds. Figure \ref{fig:barchart_distributions} shows the distribution of surface types for both training  and test sets. 
 
\begin{figure}[h!]
     \begin{center}
     \begin{tabular}{ccc}
          \includegraphics[scale=0.03, trim={0.5cm 0.10cm 0.5cm 0.5cm}, clip]{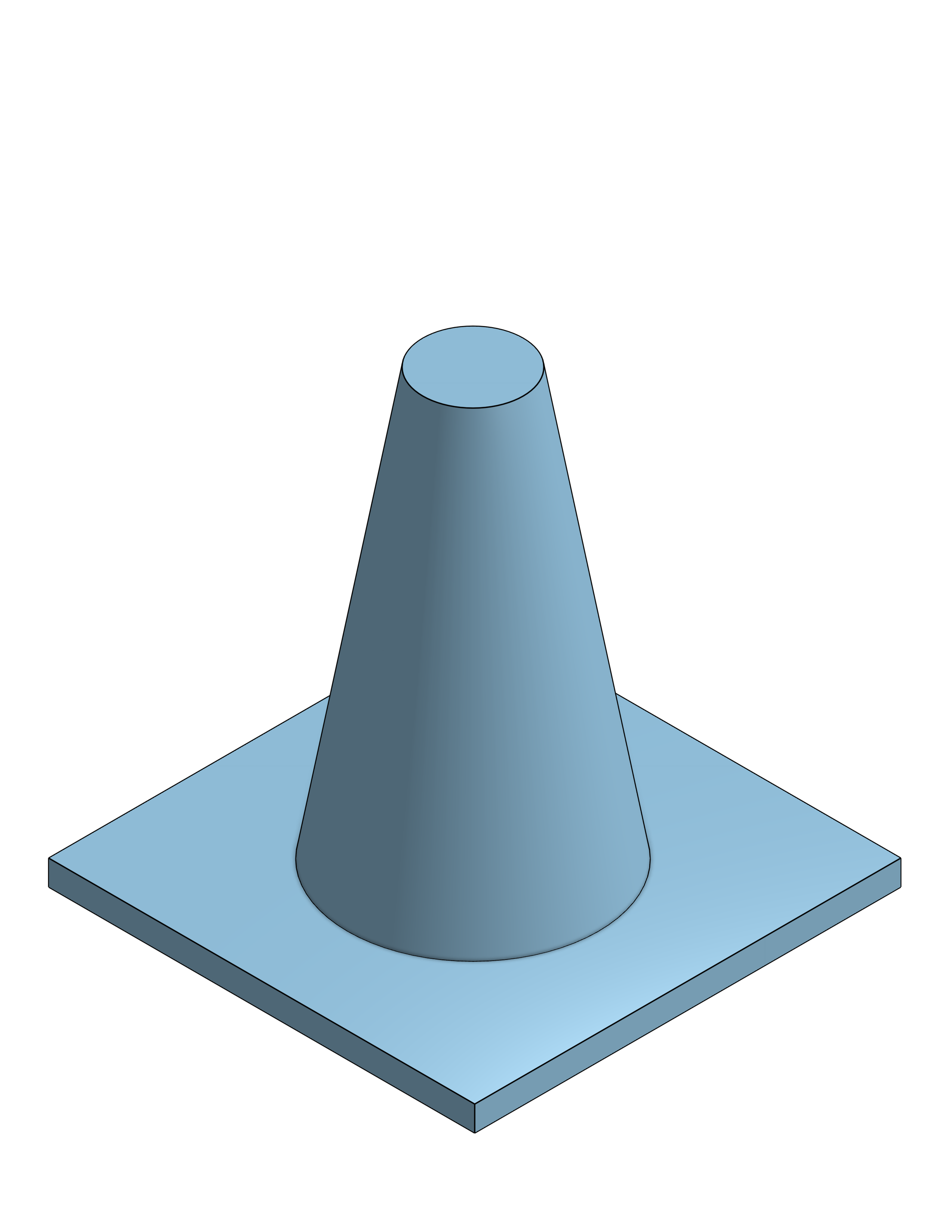}
          &
          \includegraphics[scale=0.03, trim={13cm 0.10cm 13cm 7.5cm}, clip]{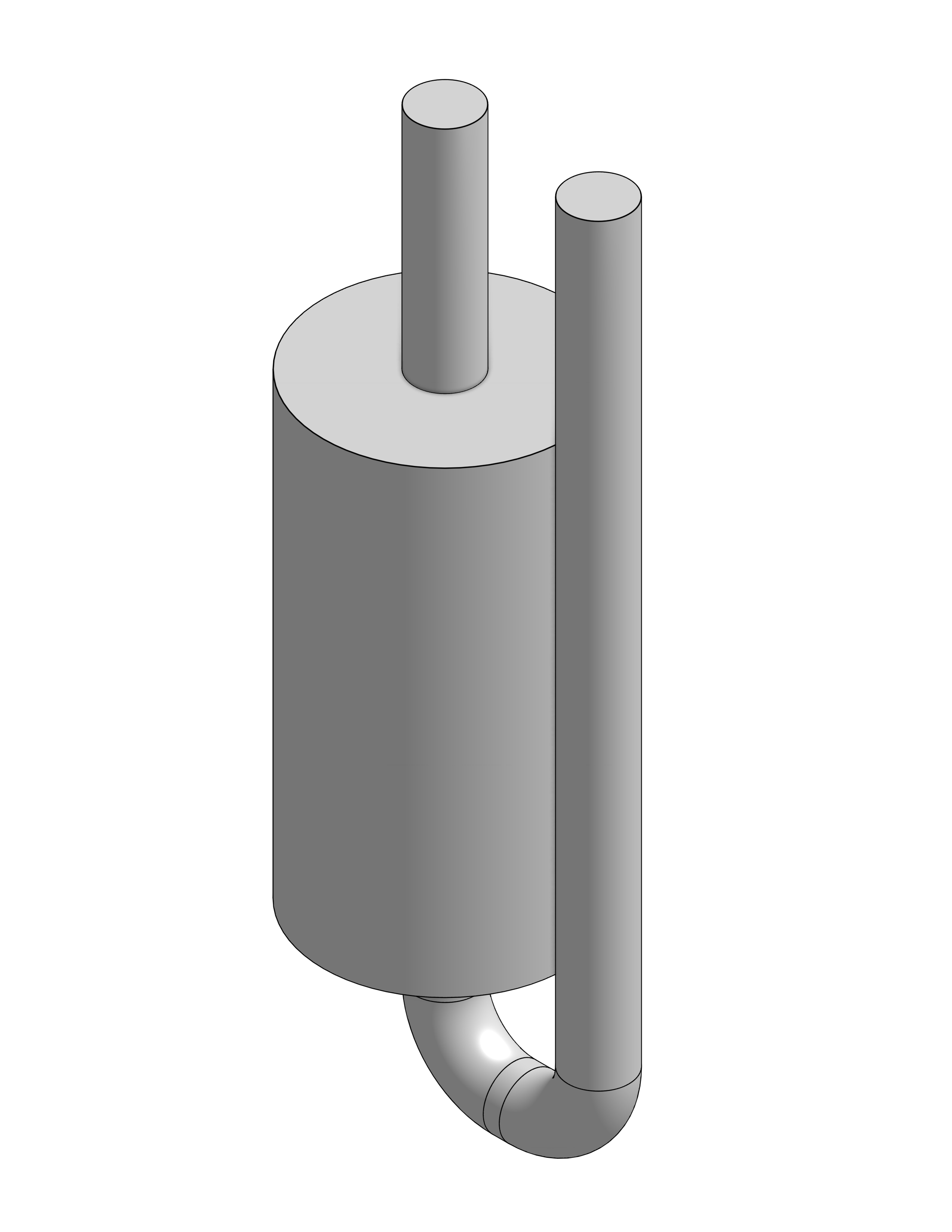}
          &
          \includegraphics[scale=0.03, trim={0.5cm 5cm 0.5cm 5cm}, clip]{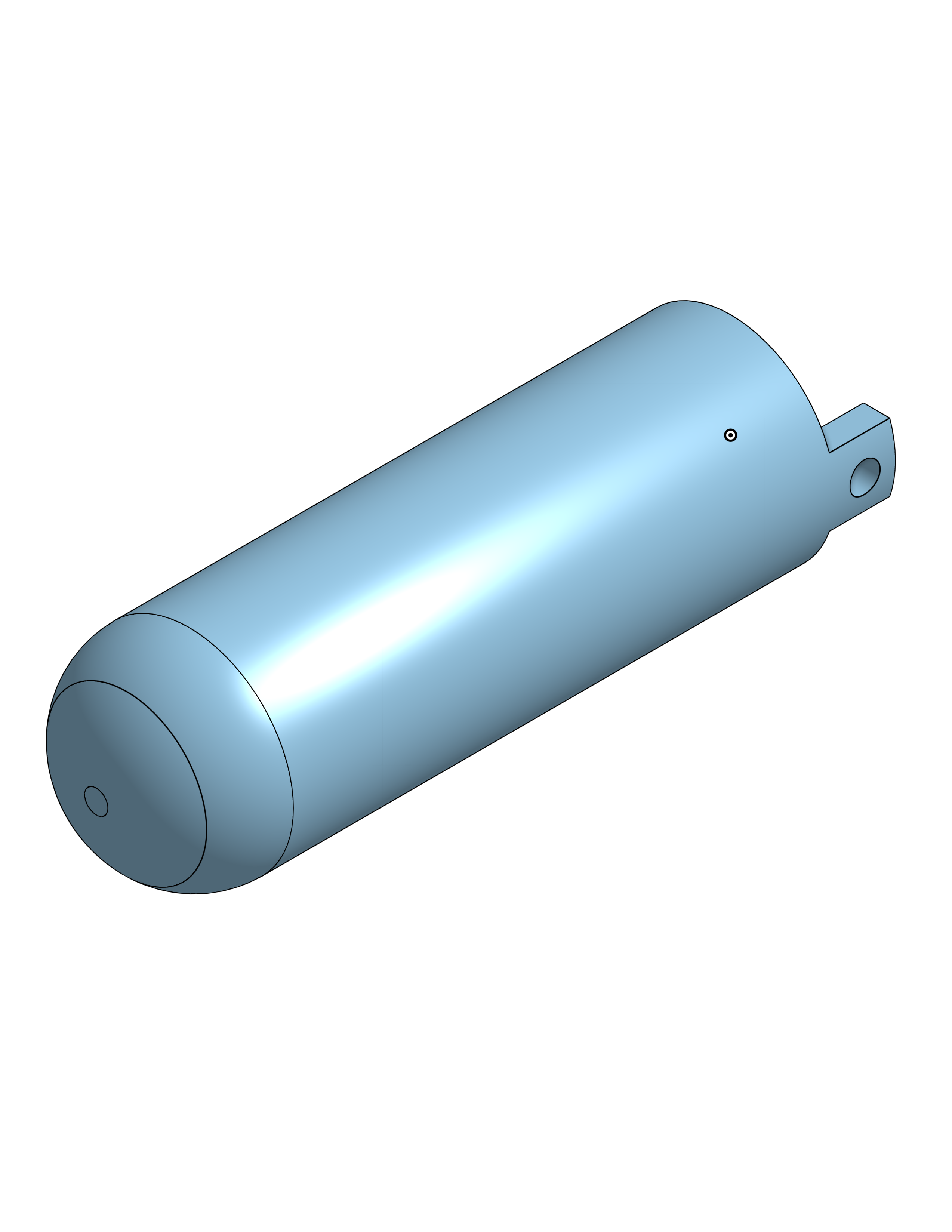}
     \end{tabular}
     \end{center}
     \caption{Example of models obtained using Onshape.}
     \label{fig:example_models}
 \end{figure}

\begin{figure}[h!]
     \begin{center}
     \resizebox{0.50\textwidth}{!}{
     \begin{tabular}{ccc}
          \includegraphics[scale=0.04, trim={7.5cm 17.5cm 0.5cm 17.5cm}, clip]{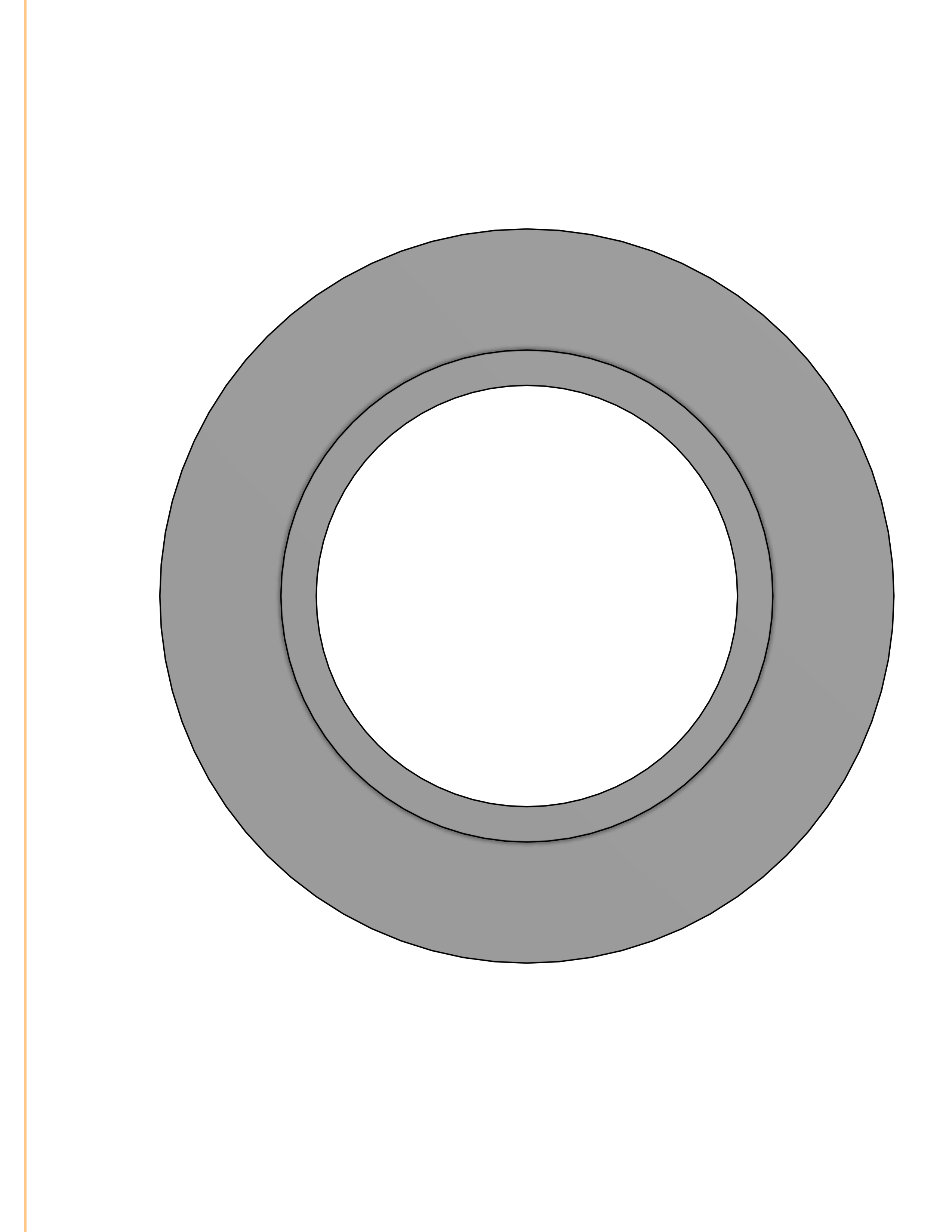}
          &
          \includegraphics[scale=0.375, trim={0cm 0cm 0cm 0cm}, clip]{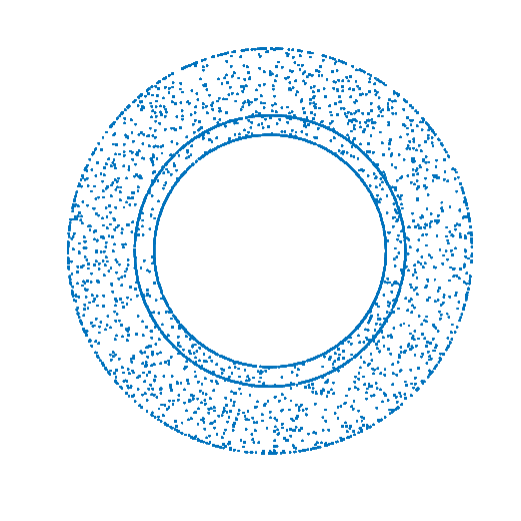}
          &
          \includegraphics[scale=0.3750, trim={0cm 0cm 0cm 0cm}, clip]{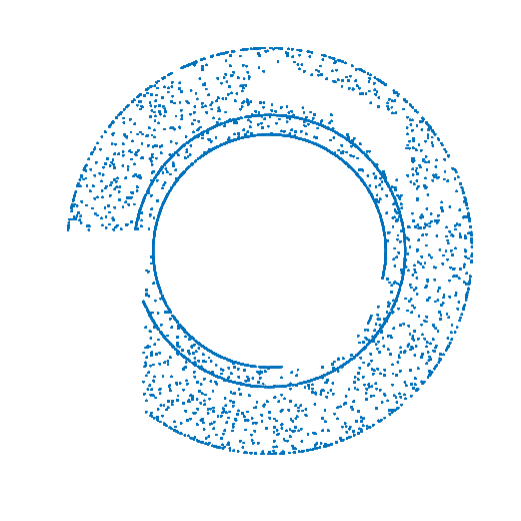}
          \\
          (a) & (b) & (c)
     \end{tabular}
     }
     \end{center}
     \caption{Example of point cloud creation. The initial object in (a) is sampled at a chosen density (b) and then perturbed by simulating missing data (c).}
     \label{fig:dataset_creation}
 \end{figure}

The dataset generation process, in the most general form, has been carried out by the following three steps:
 \begin{enumerate}
     \item \textit{Model creation}. We created part of the  models,  by using the publicly available interface hosted by Onshape, while the remaining part was collected from the ABC dataset \cite{Koch_2019}, which was derived, in turn, from the Onshape public collection. Models gathered from the ABC dataset have been filtered by manually correcting the parts presenting minimum flaws and rejecting low quality models, in order to avoid rare yet bothersome imperfections, such as overlapping or repeating patches.
    Some examples of CAD objects from Onshape are displayed in Figure \ref{fig:example_models}.
     \item \textit{Parametric and implicit representations}. The generation of B-rep models was crucial to extract the parametric representation behind each geometric primitive; in our case, the parametric representations for each patch have been obtained by processing the STEP files produced by Onshape in GMSH \cite{Geuzaine:2009}; nevertheless, we emphasize that other software could be considered too (e.g., \cite{Mathur:2020}). Several methods to compute the implicit representation from a parametric form are nowadays available. We here consider the numerical approach known as \emph{approximate implicitization}, introduced in \cite{Dokken:1997} and further delevoped in \cite{Barrowclough:2012}. One of the advantages of this approach is that it provides exact implicit representations when the exact total degree is selected; we remind that a bivariate polynomial has total degree $n$ if all monomials $x^iy^j$ are such $i+j\le{}n$, and there exists at least one monomial $x^{i}y^{j}$ such that $i+j=n$.
     \item \textit{Point cloud extraction}. CAD objects are sampled at different densities, and optionally manually postprocessed by using CloudCompare\footnote{CloudCompare (version 2.10.2), http://www.cloudcompare.org/} to simulate missing data. To give an example, Figure \ref{fig:dataset_creation}(a) shows a model from Onshape, which is then sampled and postprocessed in \ref{fig:dataset_creation}(b-c).
 \end{enumerate}

\begin{figure*}
\centering
\resizebox{0.80\textwidth}{!}{
    \centering
    \begin{tabular}{|c|c|c|c|c|c|c|}
    \hline
      \includegraphics[width=2cm]{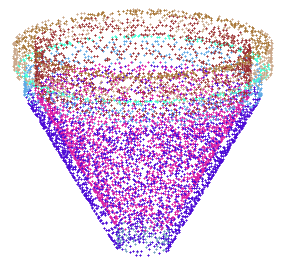}&
      \includegraphics[width=2cm]{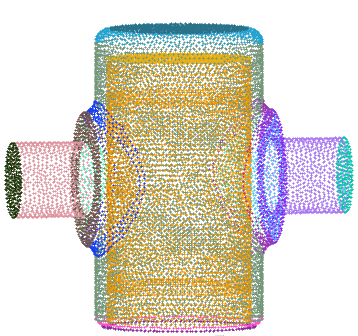}&
      \includegraphics[width=2cm]{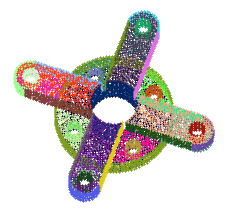}&
      \includegraphics[width=2cm]{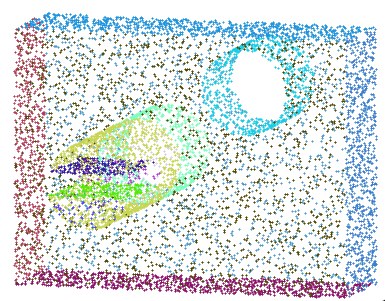}&
      \includegraphics[width=2cm]{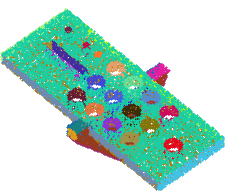}&
      \includegraphics[width=2cm]{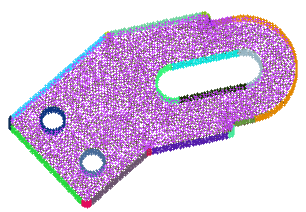}&
      \includegraphics[width=2cm]{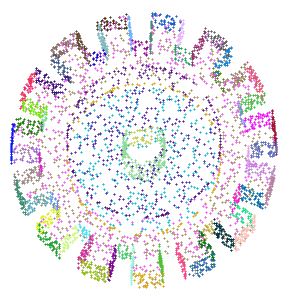}\\ 
      PC $1$ & PC $2$ & PC$3$ & PC $4$ & PC $5$ & PC$6$ & PC $7$ \\
      \hline
       \includegraphics[width=2cm]{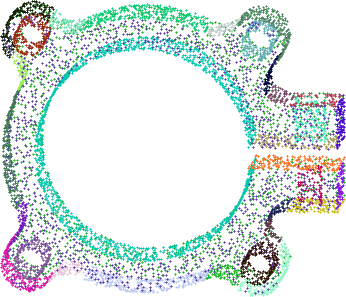}&
      \includegraphics[width=1.5cm]{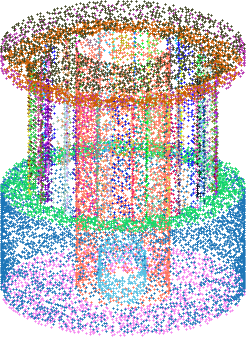}&
      \includegraphics[width=2cm]{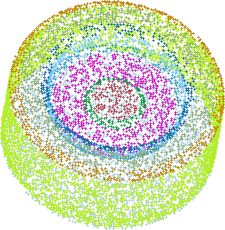}&
      \includegraphics[width=0.9cm]{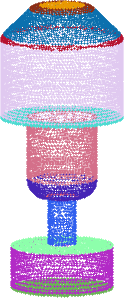}&
      \includegraphics[width=2cm]{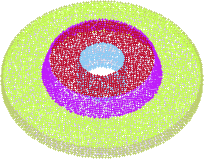}&
      \includegraphics[width=1.8cm]{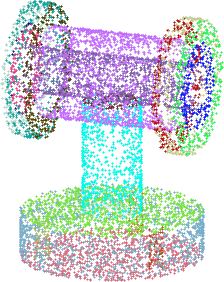}&
      \includegraphics[width=2cm]{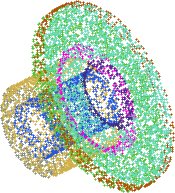}\\ 
      PC $8$ & PC $9$ & PC$10$ & PC $11$ & PC $12$ & PC$13$ & PC $14$ \\
      \hline
      \includegraphics[width=2cm]{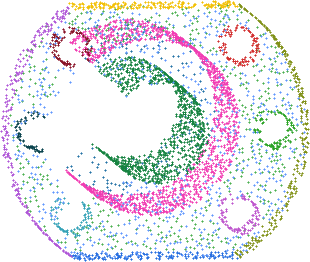}&
      \includegraphics[width=2cm]{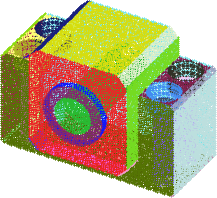}&
      \includegraphics[width=2cm]{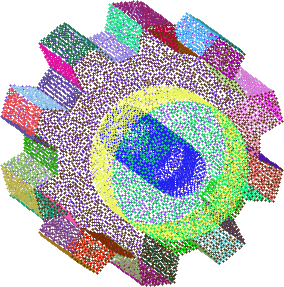}&
      \includegraphics[width=2cm]{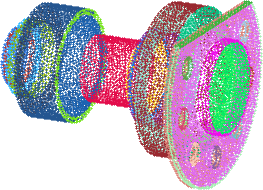}&
      \includegraphics[width=2cm]{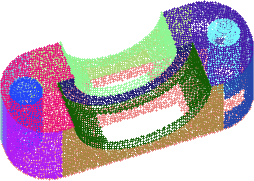}&
      \includegraphics[width=0.85cm]{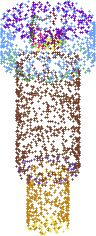}&
      \includegraphics[width=2cm]{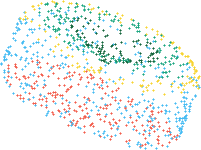}\\ 
      PC $15$ & PC $16$ & PC$17$ & PC $18$ & PC $19$ & PC$20$ & PC $21$ \\
      \hline
      \includegraphics[width=1.2cm]{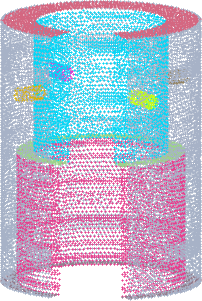}&
      \includegraphics[width=1.8cm]{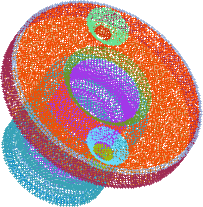}&
      \includegraphics[width=2cm]{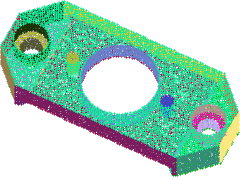}&
      \includegraphics[width=2cm]{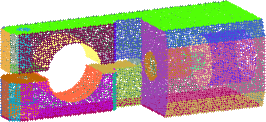}&
      \includegraphics[width=2cm]{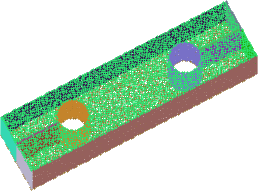}&
      \includegraphics[width=2cm]{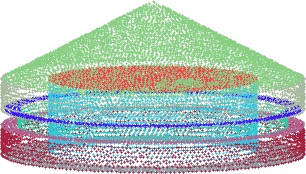}&
      \includegraphics[width=1.6cm]{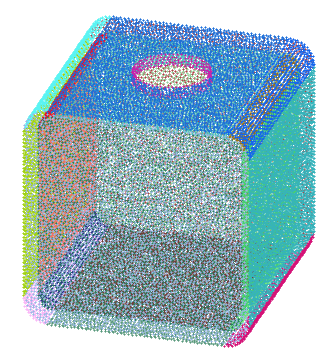}\\ 
      PC $22$ & PC $23$ & PC$24$ & PC $25$ & PC $26$ & PC$27$ & PC $28$ \\
      \hline
      \includegraphics[width=0.9cm]{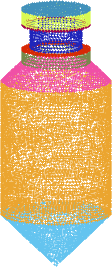}&
      \includegraphics[width=1.8cm]{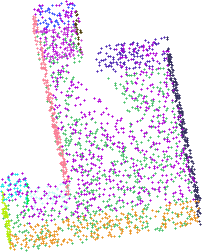}&
      \includegraphics[width=2cm]{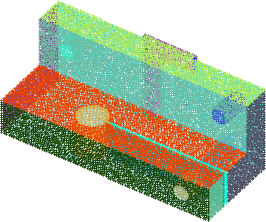}&
      \includegraphics[width=1.5cm]{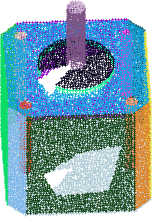}&
      \includegraphics[width=2cm]{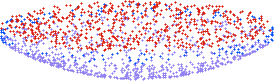}&
      \includegraphics[width=1cm]{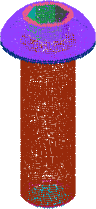}&
      \includegraphics[width=1.8cm]{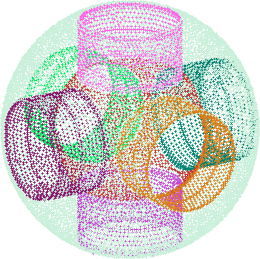}\\ 
      PC $29$ & PC $30$ & PC$31$ & PC $32$ & PC $33$ & PC$34$ & PC $35$ \\
      \hline
    \end{tabular}
    }
    \caption{The 35 point clouds used as a test set. Different colors represent different primitives, as stored in the CAD models, i.e., our ground truth.}
    \label{fig:test_set}
\end{figure*}

\subsection{Ground truth}
 \label{sec:ground}

Each model in the ground truth comes in the form of four TXT files. We here provide a description of each file content for the i-th point cloud.

\begin{description}
    \item[$PCi$] lists the three-dimensional points forming the point cloud to be segmented.
    
    \item[$PCi\_primitives$] contains the list of true primitives. For each primitive, a list of indices is provided; each index corresponds to a point in ``$PCi$", with respect to the  ordering there introduced.  For example,
    \begin{center}
    \texttt{
     Primitive6:=[4 9 184 185 186 187 188 189 190 191 192]}
    \end{center}
    means that the sixth primitive contains points number 4, 9, 184, 185, 186, 187, 188, 189, 190, 191 and 192 (where the ordering is the one in the corresponding ``$PCi$"). 
    
    \item[$PCi\_parametric$] provides, for each primitive in ``$PCi\_primitives$" corresponding to a plane, a cylinder, a cone, a sphere or a torus, its parametric representation. To give an example, 
    \begin{center}
    \texttt{
     Primitive6:=[primitive type,} \textbf{v}\texttt{]}
    \end{center}
    where  \textbf{v} is the vector that contains the parameters of the parametric representation (see \ref{sec:param_rep} for further details on the considered ordering).
    
    \item[$PCi\_implicit$] provides, for each primitive in ``$PCi\_primitives$" corresponding to a plane, a cylinder, a cone, a sphere or a torus, its implicit representation. For example,
    \begin{center}
    
     \texttt{Primitive6:=[primitive type,} \textbf{w}\texttt{]}
    \end{center}
   where  \textbf{w} is the vector that contains the coefficients of the implicit representation (see \ref{sec:impl_rep} for further details on the considered ordering).
\end{description}

The points that do not correspond to any of the simple primitives mentioned above (i.e., plane, cylinder, cone, sphere or torus) are classified as unsegmented and not explicitly reported in files $PCi\_primitives$, $PCi\_parametric$ and $PCi\_implicit$; in the original  model, these points usually originate from B-spline surfaces. We intentionally decided to insert some models with non-simple geometric primitives to check whether a candidate method can avoid misclassification.

\subsection{Quality indicators}
 \label{sec:metrics}
To evaluate the detection of simple primitives in CAD  objects, we have proposed quality measures selected  from \cite{Kuhn:2018, DezaDeza}  with particular care on what concerns their performance and approximation accuracy.

 \subsubsection{Performance measures of the point classification}
 \label{sec:measures}
 Any primitive in a model is identified by the list of points belonging to it or, equivalently, by the list of points that do not belong to it. The problem of primitive detection can therefore be easily written in terms of binary classification tasks, one per primitive in the ground truth. 
 
Let $\mathscr{P}_B$ be a a set of points in the benchmark point cloud corresponding to a specific primitive, and let $\mathscr{P}_S$ be the primitive in the segmentation to assess that most overlap with $\mathscr{P}_B$. We can define the following quantities:
 \begin{itemize}
     \item \emph{True positives}, $\text{TP}$: the number of points shared by $\mathscr{P}_B$ and $\mathscr{P}_S$.
     
     \item \emph{False positives}, $\text{FP}$: the number of points in $\mathscr{P}_S$ that do not belong to $\mathscr{P}_B$. 
     
     \item \emph{False negatives}, $\text{FN}$: the number of points in $\mathscr{P}_B$ that do not belong to $\mathscr{P}_S$.
     
    \item \emph{True negatives}, $\text{TN}$: the number of points that    do not belong to either  $\mathscr{P}_B$ nor  $\mathscr{P}_S$.
 \end{itemize}

 Based on these four quantities, we consider the following  measures:
 \begin{itemize}
     
     \item \emph{Sensitivity}, also called \emph{true positive rate}, measures the proportion of positives which are correctly identified, i.e., 
     \begin{equation*}
         \text{TPR}:=\dfrac{\text{TP}}{\text{TP}+\text{FN}}.
     \end{equation*}
    
      \emph{Specificity}, or \emph{true negative rate}, measures the proportion of true negatives that are correctly identified as such, i.e., 
     \begin{equation*}
         \text{TNR}:=\dfrac{\text{TN}}{\text{TN}+\text{FP}}.
     \end{equation*}
     
     \item \emph{Positive predictive value} is defined as the proportion of predicted positives which are actual positives, i.e., 
     \begin{equation*}
         \text{PPV}:=\dfrac{\text{TP}}{\text{TP}+\text{FP}}.
     \end{equation*}
     Similarly, \emph{negative predictive value} is given by \begin{equation*}
         \text{NPV}:=\dfrac{\text{TN}}{\text{TN}+\text{FN}}.
     \end{equation*}
     
     \item \emph{Accuracy} is the ratio of correct predictions to total predictions made, i.e., 
     \begin{equation*}
         \text{ACC}:=\dfrac{\text{TP} + \text{TN}}{\text{TP}+\text{TN}+\text{FP}+\text{FN}}.
     \end{equation*}
     
     \item \emph{Sørensen-Dice index}. It is given by
     \begin{equation*}
         \text{DSC}:=\dfrac{2|\mathscr{P}_B\cap{}\mathscr{P}_S|}{|\mathscr{P}_B| + |\mathscr{P}_S|}.
     \end{equation*}
     In case of binary classification, it is shown to be equivalent to
     \begin{equation*}
         \text{DSC}:=\dfrac{2\text{TP}}{2\text{TP}+\text{FP}+\text{FN}},
     \end{equation*}
     which is often referred to as \emph{$F_1$ score}.
     
 \end{itemize}
 
For more details, we refer the reader to \cite{Kuhn:2018}.
 \subsubsection{Approximation accuracy}
\label{sec:accuracy}

To measure the recognition accuracy of a specific primitive, we use the parametric and the implicit representations provided in ``$PCi\_implicit$" and ``$PCi\_parametric$".
Exploiting the notation provided before, let us consider a primitive $\mathscr{P}_S$ to be evaluated, and let $\mathcal{S}$ be the surface described by the corresponding parametric representation. When it comes to the parametric representation, we use the following two measures to evaluate the approximation accuracy of primitive $\mathscr{P}_S$:

\begin{itemize}
    \item \emph{Mean Fitting Error} (MFE):  \begin{equation}
\text{MFE}(\mathscr{P}_S,\mathcal{S}):=\dfrac{1}{|\mathscr{P}_S|}\sum_{\mathbf{x}\in\mathscr{P}_S}d(\mathbf{x},\mathcal{S})/l,
\label{eqn:MFE}  
\end{equation}

where $d$ is the Euclidean distance, and $l$ is the diagonal of the minimum bounding box containing $\mathscr{P}_S$.

     \item \emph{Directed Hausdorff distance}:
		$$d_{\text{dHaus}}(\mathscr{P}_S,\mathcal{S})=\max_{\mathbf{x}\in \mathscr{P}_S} \min_{\mathbf{y}\in \mathcal{S}}d(\mathbf{x},\mathbf{y}),$$ with $d$ the Euclidean distance. To make the measure independent from the primitive size, we normalize it with respect to the diagonal $l$ of the minimum bounding box containing $\mathscr{P}_S$.
	\end{itemize}
	The fitting accuracy for the implicit representation is evaluated by the following measure:
	\begin{itemize}
     \item \emph{Coefficient distance}:
     $$d_{1}(\mathbf{v},\mathbf{v'})=\|\mathbf{v}-\mathbf{v'} \|_1$$
     where $\mathbf{v}$ and $\mathbf{v'}$ are the coefficient vectors for the implicit representations of the primitives $\mathscr{P}_S$ and $\mathscr{P}_B$, respectively, and where $\|\cdot\|_1$ is the well-known $\ell^1$ norm. In order to make this measure consistent, we assume the coefficient vectors to be normalized, and the first nonzero entry to be positive (where the ordering is the one provided in \ref{sec:impl_rep}).
		
\end{itemize}
 
We refer the reader to \cite{DezaDeza} for further details.

\section{Test of the benchmark on two methods}
\label{sec:methods}
The proposed benchmark has been used to evaluate and compare two methods dealing with  primitive fitting.  
As guiding examples of how the benchmark works, we have selected two methods that are both available and representative of two classes of methods according to the taxonomy defined in \cite{Kaiser2019}: a clustering method based on a primitive growing framework (Section \ref{sec:RANSAC}) and a parametric method based on the Hough transform \cite{c1962method} (Section \ref{sec:Hough}). 
Both methods can be evaluated according to the measures described in Section \ref{sec:measures}, as they explicitly provide the list of the points that form any primitive; on the other hand, the approximation accuracy can be assessed only for methods that can provide parametric or implicit representations, in our examples the Hough-based fitting.

\subsection{PG: a discrete curvature-based method for point cloud segmentation}
\label{sec:RANSAC}
As a first approach, we present a curvature-based method based on a primitive growing framework, on the basis of the method proposed in \cite{Qie:2021} for triangle meshes; for the sake of brevity, we will often use the acronym PG as a shorthand for this method, where PG stands for ``Primitive Growing".

The method consists of two main steps: an initial region partitioning process based on high curvature detection and, then, a region refinement process based on slippage analysis, as summarized in Figure \ref{fig:curvature-based_framework}. These two steps run as follows:
\begin{itemize}
    \item \emph{Initial region partition}. Points that identify sharp edges are characterized via a point attribute called \emph{surface variation}, as introduced in \cite{Pauly2002}. Given a point $\mathbf{p}$ and $n$ neighbouring points, its surface variation is defined as
    \begin{equation*}
        \sigma_n(\mathbf{p}):=\dfrac{\lambda_1}{\lambda_1+\lambda_2+\lambda_3},
    \end{equation*}
    where $\lambda_1\le{}\lambda_2\le{}\lambda_3$ are the eigenvalues of the covariance matrix for the sample point $\mathbf{p}$ and its $n$ neighbouring points; 
    note that $\lambda_l$ measures the variability of the neighborhood of $n$ points along the direction of the corresponding eigenvector. In our experiments, $n$ is set to $15$ as it yields good results when considering the training set.
    Points on sharp edges are characterized by a high surface variation. These points are here selected by analysing a histogram of the surface variation values, by means of a threshold $\eta$: all points having surface variation above $\eta$ are labelled as sharpe edge points; for example, setting $\eta=0.8$ means that the points whose surface variation is higher than 80\% (top 20\%) are considered to belong to a sharp edge. In our implementation, the threshold $\eta$ is user-defined and taken in the interval $[0.65,0.95]$. Once sharp edges have been identified, a region growing approach is applied to compute a first coarse pre-segmentation, along the lines of what detailed in \cite{Pauly2002}: starting from a random seed point, its nearest neighbors  are progressively located; those points which does not belong to sharp edges will be labelled and used as new seed points, until all neighboring points are labelled. 
    
    \item \emph{Region refinement}. According to the ISO GPS invariance class \cite{ISO17450}, ``ideal" features can be categorized into seven invariance classes: planar, cylindrical, helical, spherical, revolute, prismatic, and complex. The seven invariance classes are here captured by local slippage analysis \cite{Gelfand:2004}. This step aims at decomposing any coarse segment $S$ from the previous step into simpler geometric parts. Given a point set $P$ of $n$ points from $S$, the slippable motions of $P$ are found as the motion vector $[\mathbf{r},\mathbf{t}]$ that, when applied to $P$, minimizes the motion along the normal direction at each point
    \begin{equation}
        \min\limits_{[\mathbf{r},\mathbf{t}]}\sum_{i=1}^n((\mathbf{r}\times{}\mathbf{x}_i+\mathbf{t})\cdot{}\mathbf{n}_i)^2,
        \label{eqn:slippage_max}
    \end{equation}
    where: $\mathbf{r}=(r_x,r_y,r_z)$ is a rotation vector around $x$, $y$, and $z$; $\mathbf{t}=(t_x,t_y,t_z)$ is a translational vector; $\mathbf{p}_i\in{}P$ are the $n$ samples, and $\mathbf{n}_i$ are their respective normals. Equation \ref{eqn:slippage_max} is a least-square problem which can be reduced to the linear system based on the covariance matrix of the second partial derivatives of the function in \ref{eqn:slippage_max} with respect to the rotation and translation parameters, see \cite{Gelfand:2004} for further details.
    Slippage analysis permits the detection of $3$-, $2$-, $1$- and $0$-slippable  motions, see Table \ref{tab:ISO_GPS_Invariance}. According to the primitives defined in the benchmark,  segments identified as prismatic, revolute and complex could be undersegmented;  
    to address this problem, the RANSAC method introduced in \cite{Schnabel2007} is applied; an example of prismatic segment requiring further processing is shown in Figure \ref{fig:PC_invariance_class}. Finally, points on sharp edges are assigned to the closest primitives.
    
    \begin{figure}
    \centering
    \includegraphics[scale=0.65]{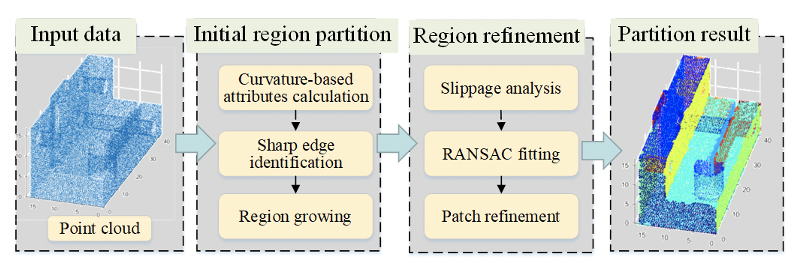}
    
    \caption{The framework of the curvature-based surface partitioning method. \label{fig:curvature-based_framework}}
    \end{figure}
    
    \begin{table}[h!]
        \centering
        \begin{tabular}{ccc}
        \toprule
             ISO GPS Invariance & Slippage & geometric primitives\\
             \bottomrule
             planar & 3 & plane \\
             spherical & 3 & sphere \\
             cylindrical & 2 & cylinder \\
             helical & 2 & - \\
             prismatic & 1 & undersegmented \\
             revolute & 1 & cone/torus \\
             complex & 0 & undersegmented \\
        \bottomrule
        \end{tabular}
        \caption{Relation between ISO GPS invariance class \cite{ISO17450} and the simple geometric primitives in this benchmark. Note that, in this terminology,
        prismatic/complex could include include planar, spherical or cylindrical primitives.}
        \label{tab:ISO_GPS_Invariance}
    \end{table}
    
    \begin{figure}[h!]
    \centering
    \includegraphics[scale=0.25, trim={0cm 2cm 0cm 2cm}, clip]{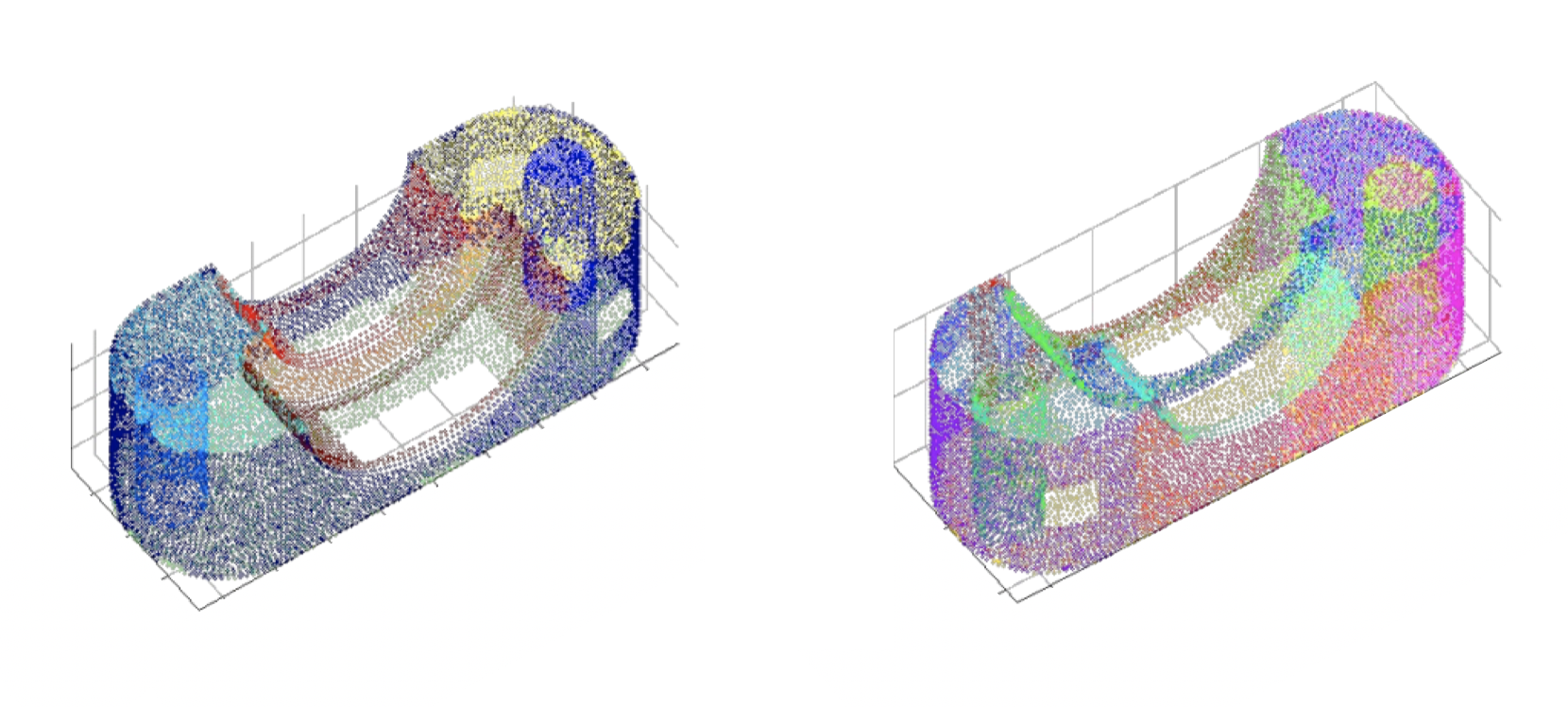}
    \caption{Example of undersegmented invariance class. On the left,  the external blue primitive is classified as prismatic. On the right, the same primitive is split into 2 planes and 2 half cylinders. \label{fig:PC_invariance_class}}
    \end{figure}

\end{itemize}

\subsubsection{Computational complexity}
The characterization of the sharp edges according to surface variation \cite{Pauly2002} is done by considering the k-nearest neighbour points with $\OOO(n\log{}n)$ operations, where $n$ represents the number of points. The growing algorithm for grouping points inside boundaries costs $\OOO(n)$ operation while the classification of each point set via slippage analysis is
$\OOO(m)$, where $m$ is the number of points in one surface portion \cite{Gelfand:2004}. The further RANSAC based segmentation in case of point sets with low slippage values (such as revolute, prismatic and complex primitives) is $\OOO(m)$ \cite{Schnabel2007}, where $m<n$ in the most common scenario. 

\subsection{HT: Simple primitive fitting based on Hough transform}
\label{sec:Hough}

\begin{figure*}[h!]
    \centering
   \includegraphics[width=17.0cm]{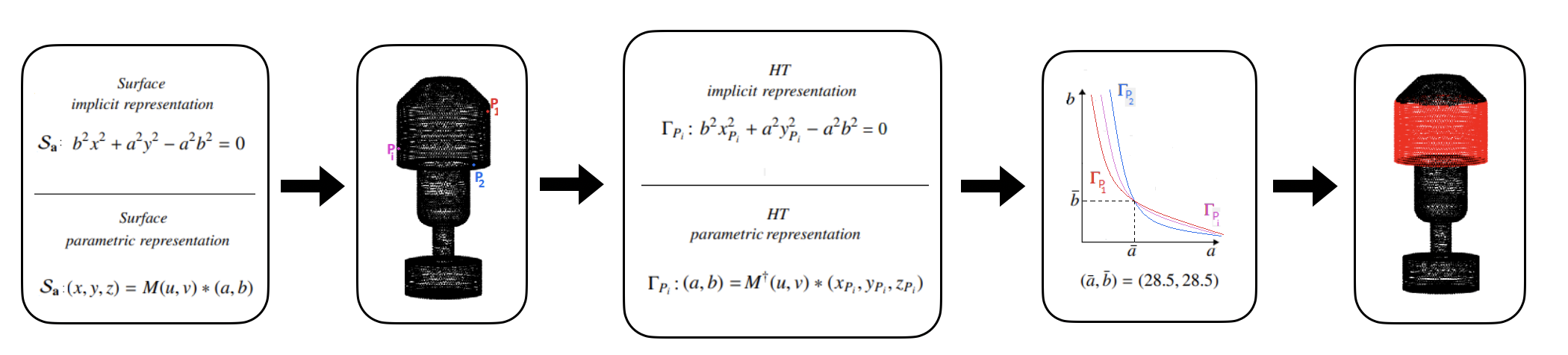}
    \caption{The HT-based paradigm: a visual illustration of how a cylinder representation $S_a$ is converted by the Hough transform into $|P_i|$ hypersurfaces $\Gamma_{P_i}$; then the intersection of the hypersurfaces $\Gamma_{P_i}$ identifies the parameters $(\bar{a},\bar{b})$ that correspond to the red cylinder in the right.}
    \label{fig:HT_framework}
\end{figure*}

In this section, we consider a method to segment and to fit a point cloud $PC$ with surface primitives using the Hough Transform (HT) technique. The general HT-framework deals with the problem of finding a surface $\mathcal{S}_\mathbf{\bar{a}}$ – within a family $\mathcal{F}=\{\mathcal{S}_\mathbf{a}\}$ of surfaces dependent on a set of parameters $\mathbf{a}=(a_1,...,a_n)$ – that best approximates a particular shape. The common strategy to identify the solution (or a solution) consists in a procedure whereby each point in $PC$ votes a $n$-uple $\mathbf{a}$ in the parameter space; the most voted $n$-uple $\mathbf{\bar{a}}$ corresponds to the most representative surface $\mathcal{S}_\mathbf{\bar{a}}$ for a dense subset of $PC$. Figure \ref{fig:HT_framework} illustrates how the Hough transform converts the problem of fitting points on a primitive into the problem of fitting the parameters of a family of primitives into points.

This method is based on the theory related to the extension of the Hough transform to general algebraic objects \cite{beltrametti2012algebraic}. This theory is very broad and can be used for many types of primitives, for instance in \cite{BELTRAMETTI2020} is used for fitting point sets with ellipsoids and can deal with non-simple primitives, such as helical surfaces. The families of primitives included in this benchmark are planes, cylinders, spheres, cones and tori.
Once a family of primitives $\mathcal{F}$ is selected, the main steps can be summarized as follows:
\begin{itemize}
    \item \textit{Inizialization and estimation of the accumulator function.} Once the family $\mathcal{F}$ is chosen, a region \textit{T} of the parameter space is selected exploiting the knowledge of the geometric characteristics of $\mathcal{F}$ (e.g., bounding box). Then, it is discretized into cells, which are uniquely identified by the coordinates of their centre. This space is associated with an accumulator function $\mathcal{H}$, discretized as a matrix. Its entries are in a one-to-one correspondence with the cells of \textit{T}. An entry of $\mathcal{H}$ is increased by $1$ each time the HT of a point $P$, $\Gamma_P$, intersects the corresponding cell. 

    \item \textit{Selection of potential fitting primitives.} In the case the input point cloud is composed of different primitives, the peaks of $\mathcal{H}$ identify the potential primitives $\mathcal{S}_\mathbf{\bar{a}_i}$ that might fit different parts $\mathcal{X}_i\subseteq PC$. Then, the cells corresponding to the peak values of the accumulator function $\mathcal{H}$ are identified by studying its \emph{topological persistence} (see \cite{EdelZomo2002}). In our implementation, the peaks that correspond to primitives are automatically recognised by keeping the local maxima with a persistence higher than 10\% of the maximum value of $\mathcal{H}$, using the algorithm for persistent maxima proposed in \cite{biasotti2016tracking}. The coordinates of the cell centres of the maxima or the peaks of the accumulator function correspond to the parameters of potentially recognised surface primitives. 
\end{itemize}

\begin{figure}
    \centering
   \begin{tabular}{|c|c|c|c|}
   \hline
  &&& \\
  \includegraphics[width=1cm]{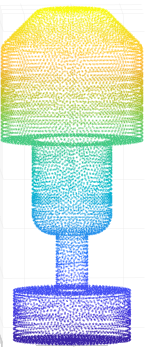}&
  \includegraphics[width=3.4cm]{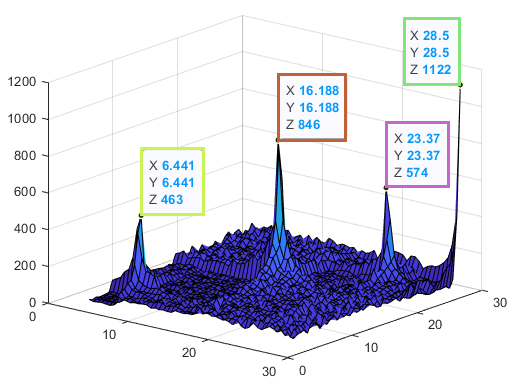} &
  \includegraphics[width=1.145cm]{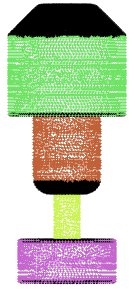}&
  \includegraphics[width=1.13cm]{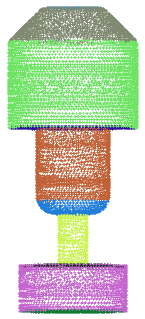}\\
  (a) & (b) & (c) & (d) \\
  \hline
\end{tabular}
    \caption{HT framework. In (a), an example of input point cloud; in (b), the accumulator function associated with the search for cylinders and the peaks found by the method for persistent maxima; in (c), the four cylinders corresponding to the four peaks (black stands for unclassified points). The final outcome, after searching for all simple geometric primitives, is shown in (d).}
    \label{fig:cilindriHT}
\end{figure}

  Since it can happen that more types of primitives fit the same dense subset $\mathcal{X}_i$ (or a part of it), the \emph{Mean Fitting Error} (see Equation \ref{eqn:MFE}) is used to evaluate the approximation accuracy of each primitive. Then, if $\mathcal{S}_\mathbf{\bar{a}_{i,1}}$ and $\mathcal{S}_\mathbf{\bar{a}_{i,2}}$ are two candidate primitives, the fitting errors $\text{MFE}(\mathcal{X}_i,\mathcal{S}_\mathbf{\bar{a}_{i,1}})$ and $\text{MFE}(\mathcal{X}_i,\mathcal{S}_\mathbf{\bar{a}_{i,2}})$ between each primitive and $\mathcal{X}_i$ are calculated; the primitive having lowest error is kept. The final result is the partitioning of the input point cloud $PC$ into several subsets in such a way that points of the same segment are well approximated by the same primitive. Figure \ref{fig:cilindriHT} summarizes the HT framework. In particular, Figure \ref{fig:cilindriHT}(b) shows an example of accumulator matrix referred to the recognition of the cylinders, while the four cylinders corresponding to the four peaks are highlighted on the original point cloud in Figure \ref{fig:cilindriHT}(c). Finally, Figure \ref{fig:cilindriHT}(d) exhibits the resulting segmentation.
   
\subsection{Evaluation}
We here analyse the performance of the methods outlined in Sections \ref{sec:RANSAC} and \ref{sec:Hough}, with the purpose of showing how the benchmark works. Firstly, we compare the quality of the segments/primitives found against a ground-truth; the measures involved do not require an explicit representation of the primitive equation, and thus can be applied to both methods. Secondly, we consider the accuracy of the parametric/implicit representations: in our case, this comes down to the analysis of the method
introduced in Section \ref{sec:Hough}.

\subsubsection*{Performance measures of the point classification}
Figures \ref{fig:res_RANSAC} and \ref{fig:res_HT} in \ref{sec:App_segmentations} provide the segmentation results obtained by the primitive growing (PG) and Hough transform (HT) based approaches, colored as follows: given a model and an approach, for each primitive in the benchmark we find the most overlapping segmented primitive; misclassified points are colored in black, while correct matches follows the $1-1$ primitive-color correspondence from Figure \ref{fig:test_set}. 

Table \ref{tab:classification_performance_metrics} summarizes the performances of the two methods over all the test set models. Each row correspond to a model; for each model, the table provides information on the number of true and predicted primitives, as well as the accuracy measures introduced in Section \ref{sec:measures}. For each metric, two columns are considered, respectively referring to the PG- and the HT-based approaches. 

\begin{table*}[tbh!]
    \centering
    \resizebox{0.97\textwidth }{!}{\begin{tabular}{|c|c|c|c|c|c|c|c|c|c|}
    \hline
    & \multirow{3}{*}{$\#$ points} & \multirow{3}{*}{\makecell{$\#$ true \\ primitives}} & $\#$ predicted & \multirow{2}{*}{DSC} &  \multirow{2}{*}{PPV} & \multirow{2}{*}{TPR} &  \multirow{2}{*}{TNR} & \multirow{2}{*}{NPV} & \multirow{2}{*}{ACC} \\
    & & & primitives &  & & & & & \\
    & & & \begin{tabular}{>{\centering}p{5mm}|>{\centering\arraybackslash}p{5mm}} PG & HT \\ \end{tabular} & \begin{tabular}{>{\centering}p{5mm}|>{\centering\arraybackslash}p{5mm}} PG & HT \\ \end{tabular} & \begin{tabular}{>{\centering}p{5mm}|>{\centering\arraybackslash}p{5mm}} PG & HT \\ \end{tabular} & \begin{tabular}{>{\centering}p{5mm}|>{\centering\arraybackslash}p{5mm}} PG & HT \\ \end{tabular} & \begin{tabular}{>{\centering}p{5mm}|>{\centering\arraybackslash}p{5mm}} PG & HT \\ \end{tabular} & \begin{tabular}{>{\centering}p{5mm}|>{\centering\arraybackslash}p{5mm}} PG & HT \\ \end{tabular} & \begin{tabular}{>{\centering}p{5mm}|>{\centering\arraybackslash}p{5mm}} PG & HT \\ \end{tabular}\\
    
        \hline
        PC $1$ & $7,500$ & $8$ & \begin{tabular}{>{\centering}p{5mm}|>{\centering\arraybackslash}p{5mm}} $12$ & $8$ \\ \end{tabular} & \begin{tabular}{c|c} $0.370$ &  $0.987$\\ \end{tabular} &
        \begin{tabular}{c|c}  $0.693$ &  $0.989$\\ \end{tabular} & \begin{tabular}{c|c} $0.332$ &  $0.985$\\ \end{tabular} & 
        \begin{tabular}{c|c} $0.968$ &  $0.998$ \\ \end{tabular} & \begin{tabular}{c|c} $0.696$ &  $0.995$\\ \end{tabular} & 
        \begin{tabular}{c|c} $0.706$ &  $0.995$\\ \end{tabular} \\
        \hline
         PC $2$ & $20,621$ & $17$ & \begin{tabular}{>{\centering}p{5mm}|>{\centering\arraybackslash}p{5mm}} $13$ &  $16$ \\ \end{tabular} & \begin{tabular}{c|c} $0.656$ &  $0.921$\\ \end{tabular} & \begin{tabular}{c|c}  $0.795$ &  $0.960$\\ \end{tabular} & \begin{tabular}{c|c} $0.632$ &  $0.937$\\ \end{tabular} & \begin{tabular}{c|c} $0.994$ &  $0.999$ \\ \end{tabular} & \begin{tabular}{c|c} $0.947$ &  $0.998$\\ \end{tabular} & \begin{tabular}{c|c} $0.944$ &  $0.998$\\ \end{tabular} \\
        \hline
         PC $3$ & $9,723$ & $35$ & \begin{tabular}{>{\centering}p{5mm}|>{\centering\arraybackslash}p{5mm}} $36$ &  $35$ \\ \end{tabular} & \begin{tabular}{c|c} $0.843$ &  $0.896$\\ \end{tabular} & \begin{tabular}{c|c}  $0.778$ &  $0.828$\\ \end{tabular} & \begin{tabular}{c|c} $0.943$ &  $0.986$\\ \end{tabular} & \begin{tabular}{c|c} $0.994$ &  $0.996$ \\ \end{tabular} & \begin{tabular}{c|c} $0.998$ &  $1.000$\\ \end{tabular} & \begin{tabular}{c|c} $0.992$ &  $0.996$\\ \end{tabular} \\
        \hline
        PC $4$ & $10,000$ & $15$ & \begin{tabular}{>{\centering}p{5mm}|>{\centering\arraybackslash}p{5mm}} $12$ &  $12$ \\ \end{tabular} & \begin{tabular}{c|c} $0.736$ &  $0.840$\\ \end{tabular} & 
        \begin{tabular}{c|c}  $0.831$ &  $0.981$\\ \end{tabular} & \begin{tabular}{c|c} $0.729$ &  $0.800$\\ \end{tabular} & 
        \begin{tabular}{c|c} $0.993$ &  $0.999$ \\ \end{tabular} & \begin{tabular}{c|c} $0.964$ &  $0.972$\\ \end{tabular} & 
        \begin{tabular}{c|c} $0.960$ &  $0.973$\\ \end{tabular} \\
        \hline
        PC $5$ & $20,000$ & $37$ & \begin{tabular}{>{\centering}p{5mm}|>{\centering\arraybackslash}p{5mm}} $32$ &  $34$ \\ \end{tabular} & \begin{tabular}{c|c} $0.480$ &  $0.839$\\ \end{tabular} & 
        \begin{tabular}{c|c}  $0.665$ &  $0.872$\\ \end{tabular} & \begin{tabular}{c|c} $0.544$ &  $0.901$\\ \end{tabular} & 
        \begin{tabular}{c|c} $0.990$ &  $0.999$ \\ \end{tabular} & \begin{tabular}{c|c} $0.925$ &  $0.978$\\ \end{tabular} & 
        \begin{tabular}{c|c} $0.918$ &  $0.978$\\ \end{tabular} \\
        \hline
         PC $6$ & $9,320$ & $26$ & \begin{tabular}{>{\centering}p{5mm}|>{\centering\arraybackslash}p{5mm}} $13$ &  $20$ \\ \end{tabular} & \begin{tabular}{c|c} $0.465$ &  $0.783$\\ \end{tabular} & 
        \begin{tabular}{c|c}  $0.671$ &  $0.857$\\ \end{tabular} & \begin{tabular}{c|c} $0.466$ &  $0.772$\\ \end{tabular} & 
        \begin{tabular}{c|c} $0.994$ &  $0.999$ \\ \end{tabular} & \begin{tabular}{c|c} $0.968$ &  $0.997$\\ \end{tabular} & 
        \begin{tabular}{c|c} $0.964$ &  $0.996$\\ \end{tabular} \\
        \hline
         PC $7$ & $5,000$ & $68$ & \begin{tabular}{>{\centering}p{5mm}|>{\centering\arraybackslash}p{5mm}} $41$ &  $69$ \\ \end{tabular} & \begin{tabular}{c|c} $0.465$ &  $0.923$\\ \end{tabular} & 
        \begin{tabular}{c|c}  $0.642$ &  $0.871$\\ \end{tabular} & \begin{tabular}{c|c} $0.429$ &  $0.994$\\ \end{tabular} & 
        \begin{tabular}{c|c} $0.993$ &  $0.999$ \\ \end{tabular} & \begin{tabular}{c|c} $0.978$ &  $1.000$\\ \end{tabular} & 
        \begin{tabular}{c|c} $0.972$ &  $0.999$\\ \end{tabular} \\
        \hline
         PC $8$ & $7,500$ & $32$ & \begin{tabular}{>{\centering}p{5mm}|>{\centering\arraybackslash}p{5mm}} $28$ &  $30$ \\ \end{tabular} & \begin{tabular}{c|c} $0.390$ &  $0.890$\\ \end{tabular} & 
        \begin{tabular}{c|c}  $0.528$ &  $0.896$\\ \end{tabular} & \begin{tabular}{c|c} $0.438$ &  $0.889$\\ \end{tabular} & 
        \begin{tabular}{c|c} $0.984$ &  $ 0.997$ \\ \end{tabular} & \begin{tabular}{c|c} $0.930$ &  $0.998$\\ \end{tabular} & 
        \begin{tabular}{c|c} $0.916$ &  $0.996$\\ \end{tabular} \\
        \hline
        PC $9$ & $17,000$ & $104$ & \begin{tabular}{>{\centering}p{5mm}|>{\centering\arraybackslash}p{5mm}} $43$ &  $40$ \\ \end{tabular} & \begin{tabular}{c|c} $0.403$ &  $0.568$\\ \end{tabular} & 
        \begin{tabular}{c|c}  $0.607$ &  $0.848$\\ \end{tabular} & \begin{tabular}{c|c} $0.335$ &  $0.456$\\ \end{tabular} & 
        \begin{tabular}{c|c} $ 0.998$ &  $0.999$ \\ \end{tabular} & \begin{tabular}{c|c} $0.993$ &  $0.996$\\ \end{tabular} & 
        \begin{tabular}{c|c} $0.991$ &  $0.995$\\ \end{tabular} \\
        \hline
        PC $10$ & $10,000$ & $10$ & \begin{tabular}{>{\centering}p{5mm}|>{\centering\arraybackslash}p{5mm}} $7$ &  $10$ \\ \end{tabular} & \begin{tabular}{c|c} $0.627$ &  $0.919$\\ \end{tabular} & 
        \begin{tabular}{c|c}  $0.793$ &  $0.926$\\ \end{tabular} & \begin{tabular}{c|c} $0.555$ &  $ 0.938$\\ \end{tabular} & 
        \begin{tabular}{c|c} $0.988$ &  $0.995$ \\ \end{tabular} & \begin{tabular}{c|c} $0.949$ &  $0.996$\\ \end{tabular} & 
        \begin{tabular}{c|c} $0.943$ &  $ 0.993$\\ \end{tabular} \\
        \hline
        PC $11$ & $13,201$ & $13$ & \begin{tabular}{>{\centering}p{5mm}|>{\centering\arraybackslash}p{5mm}} $9$ &  $10$ \\ \end{tabular} & \begin{tabular}{c|c} $0.705$ &  $0.805$\\ \end{tabular} & 
        \begin{tabular}{c|c}  $0.874$ &  $0.959$\\ \end{tabular} & \begin{tabular}{c|c} $0.677$ &  $ 0.770$\\ \end{tabular} & 
        \begin{tabular}{c|c} $0.993$ &  $0.998$ \\ \end{tabular} & \begin{tabular}{c|c} $0.937$ &  $0.973$\\ \end{tabular} & 
        \begin{tabular}{c|c} $0.934$ &  $0.973$\\ \end{tabular} \\
        \hline
        PC $12$ & $12,327$ & $6$ & \begin{tabular}{>{\centering}p{5mm}|>{\centering\arraybackslash}p{5mm}} $6$ &  $6$ \\ \end{tabular} & \begin{tabular}{c|c} $0.894$ &  $0.998$\\ \end{tabular} & 
        \begin{tabular}{c|c}  $0.852$ &  $0.997$\\ \end{tabular} & \begin{tabular}{c|c} $ 0.982$ &  $1.000$\\ \end{tabular} & 
        \begin{tabular}{c|c} $0.984$ &  $0.999$ \\ \end{tabular} & \begin{tabular}{c|c} $0.994$ &  $1.000$\\ \end{tabular} & 
        \begin{tabular}{c|c} $0.981$ &  $0.999$\\ \end{tabular} \\
        \hline
        PC $13$ & $10,000$ & $21$ & \begin{tabular}{>{\centering}p{5mm}|>{\centering\arraybackslash}p{5mm}} $17$ &  $16$ \\ \end{tabular} & \begin{tabular}{c|c} $0.582$ &  $0.800$\\ \end{tabular} & 
        \begin{tabular}{c|c}  $0.792$ &  $0.991$\\ \end{tabular} & \begin{tabular}{c|c} $ 0.576$ &  $0.761$\\ \end{tabular} & 
        \begin{tabular}{c|c} $0.993$ &  $1.000$ \\ \end{tabular} & \begin{tabular}{c|c} $0.951$ &  $0.983$\\ \end{tabular} & 
        \begin{tabular}{c|c} $0.946$ &  $0.984$\\ \end{tabular} \\
        \hline
        PC $14$ & $7,500$ & $8$ & \begin{tabular}{>{\centering}p{5mm}|>{\centering\arraybackslash}p{5mm}} $6$ &  $8$ \\ \end{tabular} & \begin{tabular}{c|c} $0.623$ &  $0.963$\\ \end{tabular} & 
        \begin{tabular}{c|c}  $0.836$ &  $0.997$\\ \end{tabular} & \begin{tabular}{c|c} $  0.606$ &  $0.933$\\ \end{tabular} & 
        \begin{tabular}{c|c} $0.987$ &  $0.999$ \\ \end{tabular} & \begin{tabular}{c|c} $0.878$ &  $0.990$\\ \end{tabular} & 
        \begin{tabular}{c|c} $0.878$ &  $0.991$\\ \end{tabular} \\
        \hline
         PC $15$ & $5,000$ & $15$ & \begin{tabular}{>{\centering}p{5mm}|>{\centering\arraybackslash}p{5mm}} $15$ &  $14$ \\ \end{tabular} & \begin{tabular}{c|c} $0.533$ &  $0.941$\\ \end{tabular} & 
        \begin{tabular}{c|c}  $0.615$ &  $0.972$\\ \end{tabular} & \begin{tabular}{c|c} $0.566$ &  $0.933$\\ \end{tabular} & 
        \begin{tabular}{c|c} $0.981$ &  $0.999$ \\ \end{tabular} & \begin{tabular}{c|c} $0.951$ &  $0.993$\\ \end{tabular} & 
        \begin{tabular}{c|c} $0.937$ &  $0.992$\\ \end{tabular} \\
        \hline
         PC $16$ & $29,641$ & $35$ & \begin{tabular}{>{\centering}p{5mm}|>{\centering\arraybackslash}p{5mm}} $28$ &  $31$ \\ \end{tabular} & \begin{tabular}{c|c} $0.676$ &  $0.848$\\ \end{tabular} & 
        \begin{tabular}{c|c}  $0.771$ &  $0.842$\\ \end{tabular} & \begin{tabular}{c|c} $0.702$ &  $0.891$\\ \end{tabular} & 
        \begin{tabular}{c|c} $0.997$ &  $0.999$ \\ \end{tabular} & \begin{tabular}{c|c} $0.986$ &  $0.994$\\ \end{tabular} & 
        \begin{tabular}{c|c} $0.984$ &  $0.993$\\ \end{tabular} \\
        \hline
        PC $17$ & $21,137$ & $45$ & \begin{tabular}{>{\centering}p{5mm}|>{\centering\arraybackslash}p{5mm}} $45$ &  $45$ \\ \end{tabular} & \begin{tabular}{c|c} $0.916$ &  $0.935$\\ \end{tabular} & 
        \begin{tabular}{c|c}  $0.859$ &  $0.889$\\ \end{tabular} & \begin{tabular}{c|c} $0.983$ &  $0.996$\\ \end{tabular} & 
        \begin{tabular}{c|c} $0.997$ &  $0.998$ \\ \end{tabular} & \begin{tabular}{c|c} $1.000$ &  $1.000$\\ \end{tabular} & 
        \begin{tabular}{c|c} $0.997$ &  $0.998$\\ \end{tabular} \\
        \hline
         PC $18$ & $16,406$ & $29$ & \begin{tabular}{>{\centering}p{5mm}|>{\centering\arraybackslash}p{5mm}} $20$ &  $29$ \\ \end{tabular} & \begin{tabular}{c|c} $0.555$ &  $0.933$\\ \end{tabular} & 
        \begin{tabular}{c|c}  $0.847$ &  $0.912$\\ \end{tabular} & \begin{tabular}{c|c} $0.536$ &  $0.978$\\ \end{tabular} & 
        \begin{tabular}{c|c} $0.994$ &  $0.999$ \\ \end{tabular} & \begin{tabular}{c|c} $0.898$ &  $0.999$\\ \end{tabular} & 
        \begin{tabular}{c|c} $0.896$ &  $0.998$\\ \end{tabular} \\
        \hline
        PC $19$ & $16,740$ & $16$ & \begin{tabular}{>{\centering}p{5mm}|>{\centering\arraybackslash}p{5mm}} $14$ &  $11$ \\ \end{tabular} & \begin{tabular}{c|c} $0.751$ &  $0.747$\\ \end{tabular} & 
        \begin{tabular}{c|c}  $0.869$ &  $0.967$\\ \end{tabular} & \begin{tabular}{c|c} $0.747$ &  $0.681$\\ \end{tabular} & 
        \begin{tabular}{c|c} $0.990$ &  $0.998$ \\ \end{tabular} & \begin{tabular}{c|c} $0.971$ &  $0.960$\\ \end{tabular} & 
        \begin{tabular}{c|c} $0.963$ &  $0.960$\\ \end{tabular} \\
        \hline
        PC $20$ & $2,500$ & $14$ & \begin{tabular}{>{\centering}p{5mm}|>{\centering\arraybackslash}p{5mm}} $8$&  $14$ \\ \end{tabular} & \begin{tabular}{c|c} $0.525$ &  $0.954$\\ \end{tabular} & 
        \begin{tabular}{c|c}  $0.789$ &  $0.917$\\ \end{tabular} & \begin{tabular}{c|c} $0.467$ &  $1.000$\\ \end{tabular} & 
        \begin{tabular}{c|c} $0.991$ &  $0.999$ \\ \end{tabular} & \begin{tabular}{c|c} $0.916$ &  $1.000$\\ \end{tabular} & 
        \begin{tabular}{c|c} $0.914$ &  $0.999$\\ \end{tabular} \\
        \hline
         PC $21$ & $1,000$ & $5$ & \begin{tabular}{>{\centering}p{5mm}|>{\centering\arraybackslash}p{5mm}} $3$&  $5$ \\ \end{tabular} & \begin{tabular}{c|c} $0.677$ &  $0.985$\\ \end{tabular} & 
        \begin{tabular}{c|c}  $0.891$ &  $0.983$\\ \end{tabular} & \begin{tabular}{c|c} $0.588$ &  $0.988$\\ \end{tabular} & 
        \begin{tabular}{c|c} $0.977$ &  $0.997$ \\ \end{tabular} & \begin{tabular}{c|c} $0.834$ &  $ 0.998$\\ \end{tabular} & 
        \begin{tabular}{c|c} $0.845$ &  $0.996$\\ \end{tabular} \\
        \hline
         PC $22$ & $26,093$ & $10$ & \begin{tabular}{>{\centering}p{5mm}|>{\centering\arraybackslash}p{5mm}} $6$ &  $10$ \\ \end{tabular} & \begin{tabular}{c|c} $0.561$ &  $0.967$\\ \end{tabular} & 
        \begin{tabular}{c|c}  $0.751$ &  $0.949$\\ \end{tabular} & \begin{tabular}{c|c} $0.586$ &  $0.987$\\ \end{tabular} & 
        \begin{tabular}{c|c} $0.994$ &  $1.000$ \\ \end{tabular} & \begin{tabular}{c|c} $0.824$ &  $ 1.000$\\ \end{tabular} & 
        \begin{tabular}{c|c} $0.822$ &  $0.999$\\ \end{tabular} \\
        \hline
        PC $23$ & $19,088$ & $13$ & \begin{tabular}{>{\centering}p{5mm}|>{\centering\arraybackslash}p{5mm}} $14$&  $12$ \\ \end{tabular} & \begin{tabular}{c|c} $0.721$ &  $0.886$\\ \end{tabular} & 
        \begin{tabular}{c|c}  $0.699$ &  $0.878$\\ \end{tabular} & \begin{tabular}{c|c} $0.828$ &  $0.930$\\ \end{tabular} & 
        \begin{tabular}{c|c} $0.985$ &  $0.997$ \\ \end{tabular} & \begin{tabular}{c|c} $0.987$ &  $0.991$\\ \end{tabular} & 
        \begin{tabular}{c|c} $0.975$ &  $0.989$\\ \end{tabular} \\
        \hline
        PC $24$ & $13,767$ & $27$ & \begin{tabular}{>{\centering}p{5mm}|>{\centering\arraybackslash}p{5mm}} $21$&  $27$ \\ \end{tabular} & \begin{tabular}{c|c} $0.742$ &  $0.916$\\ \end{tabular} & 
        \begin{tabular}{c|c}  $0.795$ &  $0.861$\\ \end{tabular} & \begin{tabular}{c|c} $0.750$ &  $0.999$\\ \end{tabular} & 
        \begin{tabular}{c|c} $0.995$ &  $0.997$ \\ \end{tabular} & \begin{tabular}{c|c} $0.992$ &  $1.000$\\ \end{tabular} & 
        \begin{tabular}{c|c} $0.987$ &  $0.998$\\ \end{tabular} \\
        \hline
         PC $25$ & $18,331$ & $38$ & \begin{tabular}{>{\centering}p{5mm}|>{\centering\arraybackslash}p{5mm}} $26$&  $35$ \\ \end{tabular} & \begin{tabular}{c|c} $0.677$ &  $0.873$\\ \end{tabular} & 
        \begin{tabular}{c|c}  $0.841$ &  $0.862$\\ \end{tabular} & \begin{tabular}{c|c} $0.665$ &  $0.921$\\ \end{tabular} & 
        \begin{tabular}{c|c} $0.997$ &  $0.998$ \\ \end{tabular} & \begin{tabular}{c|c} $0.986$ &  $0.998$\\ \end{tabular} & 
        \begin{tabular}{c|c} $0.984$ &  $0.996$\\ \end{tabular} \\
        \hline
          PC $26$ & $17,374$ & $14$ & \begin{tabular}{>{\centering}p{5mm}|>{\centering\arraybackslash}p{5mm}} $10$&  $14$ \\ \end{tabular} & \begin{tabular}{c|c} $0.663$ &  $0.975$\\ \end{tabular} & 
        \begin{tabular}{c|c}  $0.762$ &  $ 0.953$\\ \end{tabular} & \begin{tabular}{c|c} $0.607$ &  $1.000$\\ \end{tabular} & 
        \begin{tabular}{c|c} $0.988$ &  $0.999$ \\ \end{tabular} & \begin{tabular}{c|c} $0.965$ &  $1.000$\\ \end{tabular} & 
        \begin{tabular}{c|c} $0.956$ &  $0.999$\\ \end{tabular} \\
        \hline
          PC $27$ & $19,339$ & $9$ & \begin{tabular}{>{\centering}p{5mm}|>{\centering\arraybackslash}p{5mm}} $7$&  $9$  \\ \end{tabular} & \begin{tabular}{c|c} $0.789$ &  $ 0.994$\\ \end{tabular} & 
        \begin{tabular}{c|c}  $0.860$ &  $ 0.995$\\ \end{tabular} & \begin{tabular}{c|c} $0.751$ &  $0.993$\\ \end{tabular} & 
        \begin{tabular}{c|c} $0.988$ &  $1.000$ \\ \end{tabular} & \begin{tabular}{c|c} $0.971$ &  $0.999$\\ \end{tabular} & 
        \begin{tabular}{c|c} $0.963$ &  $0.999$\\ \end{tabular} \\
        \hline
          PC $28$ & $46,364$ & $21$ & \begin{tabular}{>{\centering}p{5mm}|>{\centering\arraybackslash}p{5mm}} $15$&  $21$  \\ \end{tabular} & \begin{tabular}{c|c} $0.589$ &  $0.983$\\ \end{tabular} & 
        \begin{tabular}{c|c}  $0.776$ &  $ 0.975$\\ \end{tabular} & \begin{tabular}{c|c} $ 0.551$ &  $0.993$\\ \end{tabular} & 
        \begin{tabular}{c|c} $ 0.996$ &  $0.999$ \\ \end{tabular} & \begin{tabular}{c|c} $0.963$ &  $0.999$\\ \end{tabular} & 
        \begin{tabular}{c|c} $0.960$ &  $0.999$\\ \end{tabular} \\
        \hline
         PC $29$ & $12,753$ & $9$ & \begin{tabular}{>{\centering}p{5mm}|>{\centering\arraybackslash}p{5mm}} $7$&  $9$  \\ \end{tabular} & \begin{tabular}{c|c} $0.785$ &  $0.991$\\ \end{tabular} & 
        \begin{tabular}{c|c}  $0.852$ &  $1.000$\\ \end{tabular} & \begin{tabular}{c|c} $0.758$ &  $0.983$\\ \end{tabular} & 
        \begin{tabular}{c|c} $ 0.992$ &  $1.000$ \\ \end{tabular} & \begin{tabular}{c|c} $0.985$ &  $0.998$\\ \end{tabular} & 
        \begin{tabular}{c|c} $0.980$ &  $0.999$\\ \end{tabular} \\
        \hline
         PC $30$ & $2,500$ & $13$ & \begin{tabular}{>{\centering}p{5mm}|>{\centering\arraybackslash}p{5mm}}  $8$&  $12$  \\ \end{tabular} & \begin{tabular}{c|c} $0.468$ &  $0.873$\\ \end{tabular} & 
        \begin{tabular}{c|c}  $0.666$ &  $0.873$\\ \end{tabular} & \begin{tabular}{c|c} $0.421$ &  $0.887$\\ \end{tabular} & 
        \begin{tabular}{c|c} $0.974$ &  $0.995$ \\ \end{tabular} & \begin{tabular}{c|c} $0.863$ &  $0.978$\\ \end{tabular} & 
        \begin{tabular}{c|c} $0.855$ &  $0.974$\\ \end{tabular} \\
        \hline
         PC $31$ & $22,098$ & $22$ & \begin{tabular}{>{\centering}p{5mm}|>{\centering\arraybackslash}p{5mm}}  $18$&  $23$  \\ \end{tabular} & \begin{tabular}{c|c} $0.729$ &  $0.869$\\ \end{tabular} & 
        \begin{tabular}{c|c}  $0.813$ &  $0.906$\\ \end{tabular} & \begin{tabular}{c|c} $0.758$ &  $0.887$\\ \end{tabular} & 
        \begin{tabular}{c|c} $0.992$ &  $0.995$ \\ \end{tabular} & \begin{tabular}{c|c} $0.980$ &  $0.991$\\ \end{tabular} & 
        \begin{tabular}{c|c} $0.972$ &  $0.986$\\ \end{tabular} \\
        \hline
        PC $32$ & $18,950$ & $22$ & \begin{tabular}{>{\centering}p{5mm}|>{\centering\arraybackslash}p{5mm}}  $18$&  $22$  \\ \end{tabular} & \begin{tabular}{c|c} $0.682$ &  $0.972$\\ \end{tabular} & 
        \begin{tabular}{c|c}  $0.853$ &  $0.948$\\ \end{tabular} & \begin{tabular}{c|c} $0.749$ &  $1.000$\\ \end{tabular} & 
        \begin{tabular}{c|c} $0.994$ &  $0.998$ \\ \end{tabular} & \begin{tabular}{c|c} $0.992$ &  $1.000$\\ \end{tabular} & 
        \begin{tabular}{c|c} $0.987$ &  $0.998$\\ \end{tabular} \\
        \hline
         PC $33$ & $1,500$ & $3$ & \begin{tabular}{>{\centering}p{5mm}|>{\centering\arraybackslash}p{5mm}}  $3$&  $3$   \\ \end{tabular} & \begin{tabular}{c|c} $0.811$ &  $0.978$\\ \end{tabular} & 
        \begin{tabular}{c|c}  $0.875$ &  $0.983$\\ \end{tabular} & \begin{tabular}{c|c} $0.819$ &  $0.974$\\ \end{tabular} & 
        \begin{tabular}{c|c} $ 0.910$ &  $0.983$ \\ \end{tabular} & \begin{tabular}{c|c} $0.948$ &  $0.997$\\ \end{tabular} & 
        \begin{tabular}{c|c} $0.901$ &  $0.988$\\ \end{tabular} \\
        \hline
         PC $34$ & $12,089$ & $14$ & \begin{tabular}{>{\centering}p{5mm}|>{\centering\arraybackslash}p{5mm}}  $13$&  $14$  \\ \end{tabular} & \begin{tabular}{c|c} $0.788$ &  $0.939$\\ \end{tabular} & 
        \begin{tabular}{c|c}  $0.823$ &  $0.892$\\ \end{tabular} & \begin{tabular}{c|c} $0.806$ &  $0.998$\\ \end{tabular} & 
        \begin{tabular}{c|c} $0.994$ &  $0.998$ \\ \end{tabular} & \begin{tabular}{c|c} $0.979$ &  $0.999$\\ \end{tabular} & 
        \begin{tabular}{c|c} $0.976$ &  $0.998$\\ \end{tabular} \\
        \hline
         PC $35$ & $24,068$ & $8$ & \begin{tabular}{>{\centering}p{5mm}|>{\centering\arraybackslash}p{5mm}}  $15$&  $8$  \\ \end{tabular} & \begin{tabular}{c|c} $0.866$ &  $0.975$\\ \end{tabular} & 
        \begin{tabular}{c|c}  $0.872$ &  $0.958$\\ \end{tabular} & \begin{tabular}{c|c} $0.960$ &  $0.999$\\ \end{tabular} & 
        \begin{tabular}{c|c} $0.983$ &  $0.992$ \\ \end{tabular} & \begin{tabular}{c|c} $ 0.997$ &  $1.000$\\ \end{tabular} & 
        \begin{tabular}{c|c} $0.984$ &  $0.995$\\ \end{tabular} \\
        \hline
    \end{tabular}}
    \caption{Number of fitted primitives and classification performance metrics: comparison between the PG-based and the HT-based algorithms. \label{tab:classification_performance_metrics}}
\end{table*}

\begin{figure*}[htb!]
     \begin{center}
          \includegraphics[scale=0.55, trim={2.5cm 0.10cm 2.5cm 0cm}, clip]{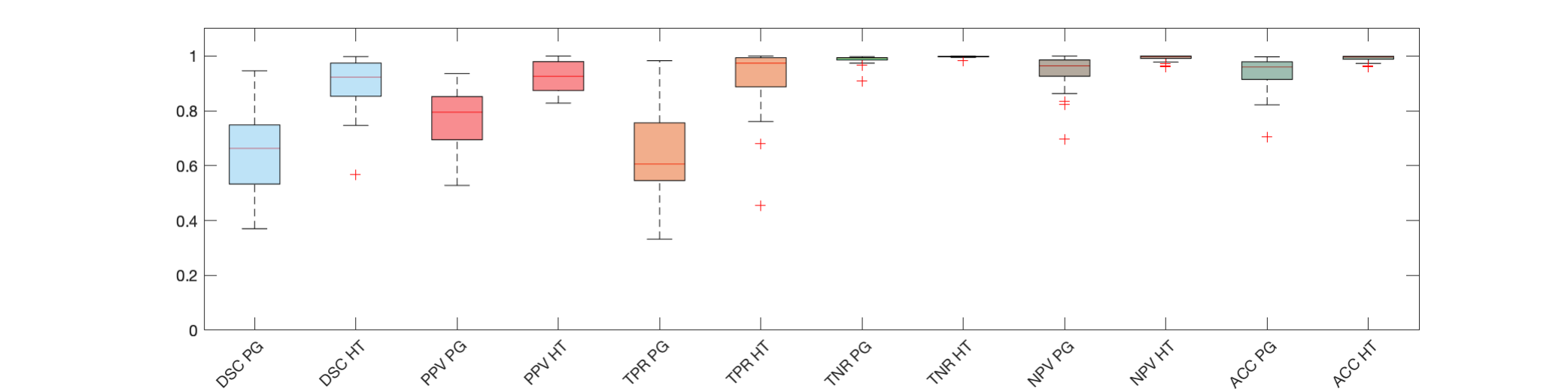}
     \end{center}
     \caption{Boxplot for the classification metrics presented in Table \ref{tab:classification_performance_metrics}. All $35$ models are here considered. \label{fig:boxplot_class}}
 \end{figure*}

\begin{figure*}[htb!]
    \begin{center}
    \begin{tabular}{c}
    \includegraphics[scale=0.55, trim={2.5cm 0.10cm 2.5cm 0cm}, clip]{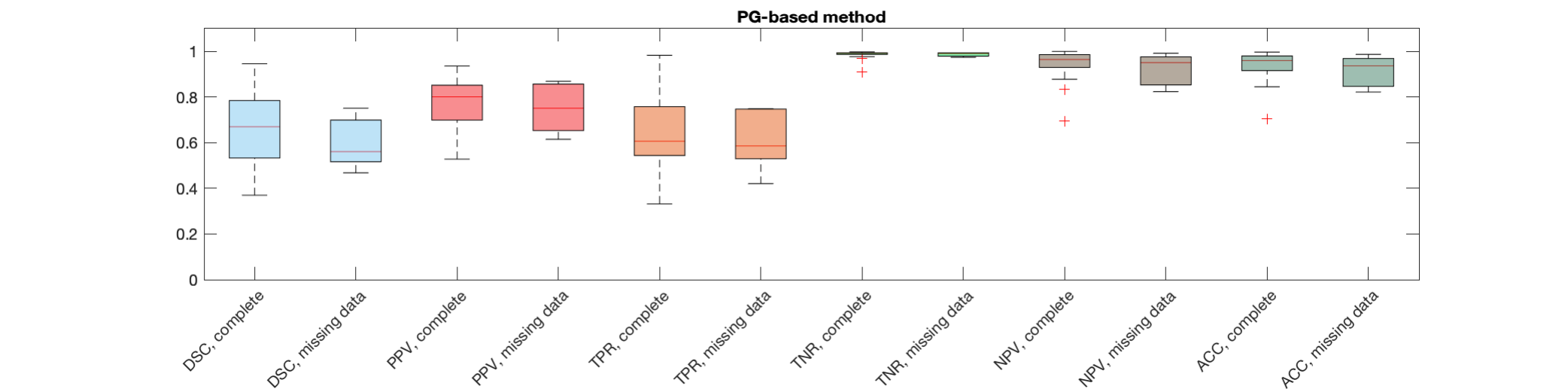}
    \\
    \includegraphics[scale=0.55, trim={2.5cm 0.10cm 2.5cm 0cm}, clip]{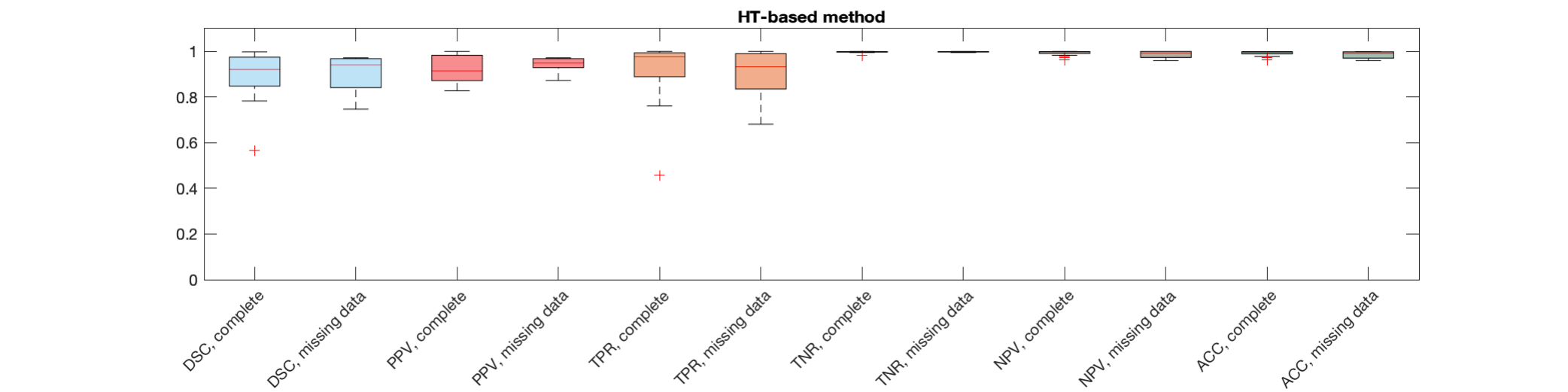}
    \end{tabular}
    \end{center}
    \caption{Performance of the PG- and HT-based methods, with an eye on models suffering from missing data. For these boxplots, we have made use of classification metrics presented in Table \ref{tab:classification_performance_metrics}.\label{fig:boxplot_RANSACHOUGH_class_missingdata}}
 \end{figure*}

To ease the analysis, the metrics are studied via boxplots:
\begin{itemize}
    \item Figure \ref{fig:boxplot_class} compares the two methods over the whole test set. A first observation of this analysis is that accuracy measures from the HT-approach have generally a lower variability. At a closer look, one can notice that the quartiles, as well as the minimum and maximum, always assume higher values when it comes to the HT-based method; in particular, the second quartile (i.e., the median) is always above $90\%$. DSC, TPR and PPV are the three accuracy measures that varies the most; this highlights that the two methods have lower performances in identifying the true positives, compared to true negatives. Both methods exhibit outliers in most of the boxplots.
    
    \item Robustness to missing data is analysed in Figure \ref{fig:boxplot_RANSACHOUGH_class_missingdata}. The HT-based method turns out to be hardly affected by such perturbation, as the inter-quartile range and the whiskers do not significantly vary; the only noteworthy variation is that of TPR, which points out a slightly decreased capability in correctly identifying positives. A more prominent variation can be noted for the PG-method.
\end{itemize}

Interestingly enough, both methods rarely suffer from oversegmentation, while it is more likely for them to undersegment. The most dramatic undersegmentation is that of point cloud $9$ (i.e., PC $9$ in Table \ref{tab:classification_performance_metrics}), where the PG-based and the HT-based methods only manage to detect $43$ and $40$ primitives, respectively, out of the $104$ there expected; this highlights possible issues when the original model has thin or small primitives.

 \subsubsection*{Approximation accuracy}
Table \ref{tab:misureParamImpl} reports the performance of the HT-based method evaluated according to the metrics reported in Section \ref{sec:accuracy}. Each row corresponds to a segmented point cloud from Figure \ref{fig:res_HT}; each column represents a different accuracy measure; for each point cloud, each measure has been obtained by averaging over all segments.

\begin{itemize}
    \item Being the MFE normalized by definition, its value can be interpreted as a percentage. From the numbers provided in the table, we can conclude that the MFE ranges from a minimum of $0.1\%$ to a maximum of $1.0\%$.
    \item The directed Hausdorff distance, in its normalized version, ranges from $0.2\%$ to $1.9\%$. The generally higher values, compared to those from the MFE, can be explained by the Hausdorff's sensitivity to outliers.
    \item The coefficient distance seems to provide a much more fluid situation. By checking the model corresponding to the highest error, we can conclude that the HT-based method has lower precision when applied to point clouds containing tori.
\end{itemize}

\begin{table}[thb!]
    \centering
    \caption{Approximation accuracy of the HT method.}
    \resizebox{0.25\textwidth}{!}{
    \begin{tabular}{|c|c|c|c|c|}
    \hline
       & MFE & $d_{\text{dHaus}}$  &$d_1$\\
       \hline
        PC $1$ & $0.002$ & $0.005$ & $0.008$ \\ 
        \hline
         PC $2$ & $0.007$ & $0.011$  &$0.149$ \\ 
        \hline
        PC $3$ & $0.002$  &  $0.004$  & $0.621$\\ 
        \hline
        PC $4$ & $0.002$ & $0.003$ &$0.000$\\ 
        \hline
         PC $5$ & $0.004$ & $0.006$  & $0.028$ \\ 
         \hline
         PC $6$ & $0.003$ & $0.005$  &  $0.001$\\ 
         \hline
         PC $7$ & $0.007$ & $0.011$  &  $0.232$\\ 
         \hline
         PC $8$ & $0.004$ & $0.013$  & $0.001$\\ 
         \hline
         PC $9$ & $0.003$ & $0.006$ &$0.000$ \\ 
         \hline
         PC $10$ & $0.005$ & $0.019$&$1.203$\\ 
         \hline
         PC $11$ & $0.003$ & $0.006$ &$0.165$\\ 
         \hline
         PC $12$ & $0.002$ & $0.004$  &$0.058$\\ 
         \hline
         PC $13$ & $0.002$ & $0.005$  &$0.000$ \\ 
         \hline
         PC $14$ & $0.004$ & $0.006$ &$1.264$\\ 
         \hline
         PC $15$ & $0.001$ & $0.003$   &$0.000$\\ 
         \hline
          PC $16$ & $0.003$ & $0.004$   &$0.029$ \\ 
         \hline
         PC $17$ & $0.003$ & $0.007$   &$0.000$\\ 
         \hline
          PC $18$ & $0.003$ & $0.004$  &$0.324$   \\ 
         \hline
         PC $19$ & $0.003$ & $0.006$   &$0.019$\\ 
         \hline
         PC $20$ & $0.002$ & $0.003$   &$0.001$\\ 
         \hline
         PC $21$ & $0.004$ & $0.007$  &$0.566$\\ 
         \hline
          PC $22$ & $0.002$ & $0.005$   &$0.000$\\ 
         \hline
          PC $23$ & $0.006$ & $0.012$  &$0.002$\\ 
         \hline
          PC $24$ & $0.001$ & $0.003$  &$0.000$\\ 
         \hline
          PC $25$ & $0.002$ & $0.003$  & $0.001$\\ 
         \hline
         PC $26$ & $0.001$ & $0.003$  & $0.000$\\ 
         \hline
         PC $27$ & $0.003$ & $0.005$  & $0.301$\\ 
         \hline
         PC $28$ & $0.001$ & $0.002$  & $0.000$\\ 
         \hline
         PC $29$ & $0.010$ & $0.003$  & $0.119$\\ 
         \hline
          PC $30$ & $0.002$ & $0.009$  & $0.003$\\ 
         \hline
          PC $31$ & $0.002$ & $0.003$  & $0.000$\\ 
         \hline
          PC $32$ & $0.006$ & $0.007$  & $0.000$\\ 
         \hline
          PC $33$ & $0.003$ & $0.006$  & $0.000$\\ 
         \hline
          PC $34$ & $0.003$ & $0.005$  & $0.004$\\ 
         \hline
          PC $35$ & $0.004$ & $0.006$  &  $0.000 $\\ 
         \hline
    \end{tabular}
    }
    \label{tab:misureParamImpl}
\end{table}

\subsubsection*{Computational time}
All tests are performed on a desktop PC equipped with an Intel Core i9 processor (at 3.6 GHz) and a Windows 10 operating system. The routines have also been tested on a MacBook Pro equipped with  macOS Catalina (version 10.15.7). We provide here some statistics of the execution times, obtained on the desktop PC:
\begin{itemize}
    \item The PG-method has minimum, mean and maximum execution time corresponding to $1.7$, $286.0$ and $19074.0$ seconds, respectively.
    \item The HT-method has minimum, mean and maximum execution time corresponding to $2.6$, $50.7$ and $358.0$ seconds, respectively.
\end{itemize}

We observe that, for small point clouds, the PG-method is generally faster, while for big point clouds it is slower.

\section{Conclusions}
In this work we have proposed Fit4CAD, a benchmark for the evaluation and comparison of methods for fitting simple geometric primitives in point clouds representing CAD objects.
The ground truth dataset of point clouds is segmented in geometric primitives and subdivided into a training set and a test set. In addition, a set of quality metrics and two fitting methods are given.
In this work, evaluation metrics are used to quantify various performance aspects of geometric primitive fitting methods. In our intent, these metrics would assist both comparing with some methods in literature and allowing 
a parameters fine-tuning of a new method, in order to optimize it on a sufficiently large set of CAD models. 

We hope the results of our comparison will inspire the development of new methods for primitive fitting, computational time being the main bottleneck in practice. In particular, it would be interesting to have a comparison with methods that use machine learning approaches, such as \cite{Tulsiani:2017}, because the dataset has been already organized in the form of a training set and a test set.

Regarding the two tested methods, the overall quality of the fitting is satisfactory for both. A rather unexpected conclusion is that over-segmentation is quite limited for both methods, while the combination of small and large primitives is a challenging task that often leads to a significant under-segmentation, see for instance the outcome on the model PC 9.

In future, we plan to continue to expand the dataset, even if we do not aim at a large scale dataset, for example by including  more complex primitives and possibly considering specific contexts such as assembly models. Moreover, the flexibility of Fit4CAD permits the insertion of other available methods to reach a more complete view of the different typologies of approaches for geometric primitive fitting. 

The benchmark is available at \url{https://github.com/chiararomanengo/Fit4CAD}. 

\section*{Acknowledgments}
\begin{sloppypar}
This work has been developed in the CNR IMATI research activities DIT.AD004.100 and DIT.AD021.080.001. The authors thank Michela Spagnuolo for the fruitful discussions and Silvia Biasotti for her substantial contribution to the paper.
\end{sloppypar}

\bibliographystyle{ieeetr}      

\appendix
\section{File type description}
The geometrical information of the primitives are provided in the files ``$PCi\_parametric$" and ``$PCi\_implicit$". Here, we describe in detail the equations of the parametric and implicit representation that they contain. Notice that, although the simple primitives shapes considered in this benchmark are polynomial (i.e., can be written as the zero set of a bivariate polynomial), the parametric representation we provide for cylinders, cones, spheres and tori are written in terms of trigonometric functions.

\subsection{Parametric representations \label{sec:param_rep}}

\begin{itemize}

    \item Plane:
\begin{equation*}
    \begin{cases}
      x = a_1u+b_1v+c_1 \\
      y = a_2u+b_2v+c_2 \\
      z = a_3u+b_3v+c_3
    \end{cases}
\end{equation*}
The parameters for a plane are stored as follows:
\begin{center}
\texttt{[Plane, [a1 a2 a3 b1 b2 b3 c1 c2 c3]]}
\end{center}

\item Cylinder:
\begin{equation*}
    \begin{cases}
      x = a_1\cos(u)+b_1\sin(u)+c_1v+d_1 \\
      y = a_2\cos(u)+b_2\sin(u)+c_2v+d_2 \\
      z = a_3\cos(u)+b_3\sin(u)+c_3v+d_3
    \end{cases}
\end{equation*}
The parameters for a cylinder are stored as follows:
\begin{center}
\texttt{[Cylinder, [a1 a2 a3 b1 b2 b3 c1 c2 c3 d1 d2 d3]}
\end{center}

\item Cone:
\begin{equation*}
    \begin{cases}
      x = a_1\cos(u)+b_1\sin(u)+c_1v\cos(u)+d_1v\sin(u)+e_1v+f_1 \\
      y = a_2\cos(u)+b_2\sin(u)+c_2v\cos(u)+d_2v\sin(u)+e_2v+f_2 \\
      z = a_3\cos(u)+b_3\sin(u)+c_3v\cos(u)+d_3v\sin(u)+e_3v+f_3
    \end{cases}
\end{equation*}
The parameters for a cone are stored as follows:
\begin{center}
\texttt{[Cone, [a1 a2 a3 b1 b2 b3 c1 c2 c3 d1 d2 d3 e1 e2 e3 f1 f2 f3]]}
\end{center}

\item Sphere:
\begin{equation*}
    \begin{cases}
      x = a_1\cos(u)\cos(v)+b_1\sin(u)\cos(v)+c_1\sin(v)+d_1 \\
      y = a_2\cos(u)\cos(v)+b_2\sin(u)\cos(v)+c_2\sin(v)+d_2 \\
      z = a_3\cos(u)\cos(v)+b_3\sin(u)\cos(v)+c_3\sin(v)+d_3
    \end{cases}
\end{equation*}
The parameters for a sphere are stored as follows:
\begin{center}
\texttt{[Sphere, [a1 a2 a3 b1 b2 b3 c1 c2 c3 d1 d2 d3]]}
\end{center}

\item Torus:
\begin{equation*}
    \begin{cases}
      x = a_1\cos(u)+b_1\sin(u)+c_1\cos(u)\cos(v)+d_1\sin(u)\cos(v)+e_1\sin(v)+f_1 \\
      y = a_2\cos(u)+b_2\sin(u)+c_2\cos(u)\cos(v)+d_2\sin(u)\cos(v)+e_2\sin(v)+f_2 \\
      z = a_3\cos(u)+b_3\sin(u)+c_3\cos(u)\cos(v)+d_3\sin(u)\cos(v)+e_3\sin(v)+f_3 \\
    \end{cases}
\end{equation*}

The parameters for a torus are stored as follows:
\begin{center}
\texttt{[Torus, [a1 a2 a3 b1 b2 b3 c1 c2 c3 d1 d2 d3 e1 e2 e3 f1 f2 f3]]}
\end{center}

\end{itemize}

\subsection{Implicit representations \label{sec:impl_rep}}

\begin{itemize}
\item Plane:
\begin{equation*}
    ax+by+cz+d=0
\end{equation*}
The coefficients for a plane are stored as follows:
\begin{center}
\texttt{[Plane, [a b c d]]}
\end{center}

\item Cylinder:
\begin{equation*}
    ax^2+by^2+cy^2+2(dxy+exz+fyz)+2(gx+hy+iz)+l=0
\end{equation*}
The coefficients for a cylinder are stored as follows:
\begin{center}
\texttt{[Cylinder,[a b c d e f g h i l]] }
\end{center}

\item Cone:
\begin{equation*}
    ax^2+by^2+cy^2+2(dxy+exz+fyz)+2(gx+hy+iz)+l=0
\end{equation*}
The coefficients for a cone are stored as follows:
\begin{center}
\texttt{[Cone, [a b c d e f g h i l]] }
\end{center}

\item Sphere:
\begin{equation*}
    ax^2+by^2+cy^2+2(dxy+exz+fyz)+2(gx+hy+iz)+l=0
\end{equation*}
The coefficients for a sphere are stored as follows:
\begin{center}
\texttt{[Sphere, [a b c d e f g h i l]] }
\end{center}

\item Torus: the coefficients of the implicit representation are provided in the form of a polynomial of degree $4$ in $x$, $y$ and $z$; they are stored in reverse lexicographic order.
\end{itemize}

\newpage
\section{Results of the test set segmentations\label{sec:App_segmentations}}

\begin{figure}[h!]
\centering
\resizebox{0.80\textwidth}{!}{
    \centering
    \begin{tabular}{|c|c|c|c|c|c|c|}
    \hline
      \includegraphics[width=2cm]{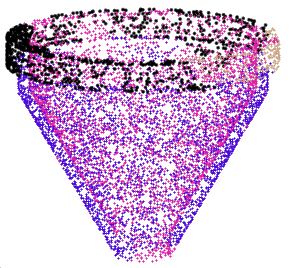}&
      \includegraphics[width=2cm]{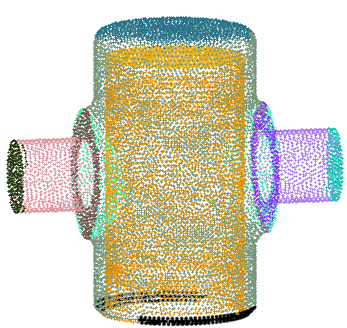}&
      \includegraphics[width=2cm]{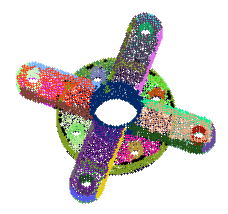}&
      \includegraphics[width=2cm]{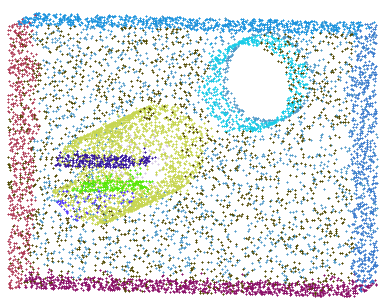}&
      \includegraphics[width=2cm]{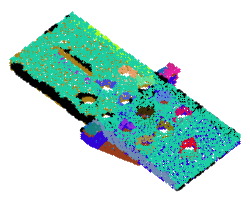}&
      \includegraphics[width=2cm]{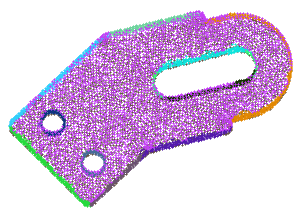}&
      \includegraphics[width=2cm]{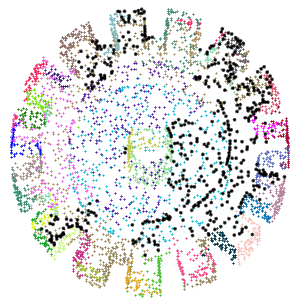}\\ 
      PC $1$ & PC $2$ & PC$3$ & PC $4$ & PC $5$ & PC$6$ & PC $7$ \\
      \hline
       \includegraphics[width=2cm]{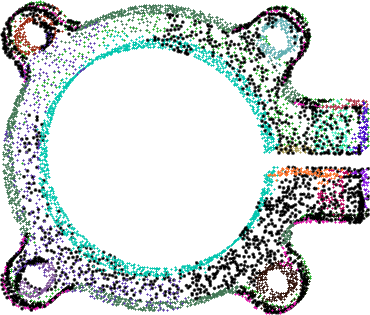}&
      \includegraphics[width=1.5cm]{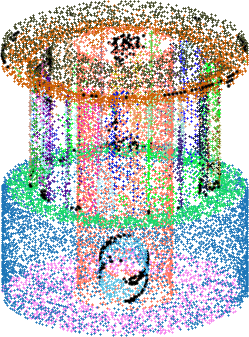}&
      \includegraphics[width=2cm]{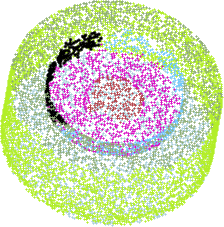}&
      \includegraphics[width=0.8cm]{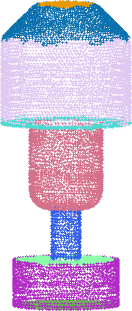}&
      \includegraphics[width=2cm]{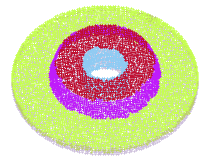}& 
      \includegraphics[width=1.8cm]{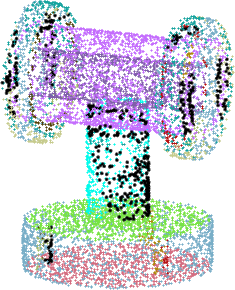}&
      \includegraphics[width=2cm]{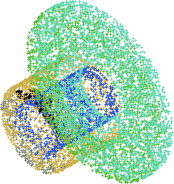}\\ 
      PC $8$ & PC $9$ & PC$10$ & PC $11$ & PC $12$ & PC$13$ & PC $14$ \\
      \hline
      \includegraphics[width=2cm]{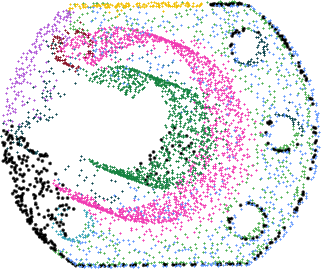}&
      \includegraphics[width=2cm]{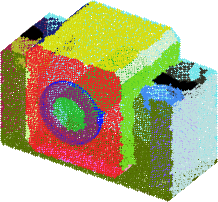}&
      \includegraphics[width=2cm]{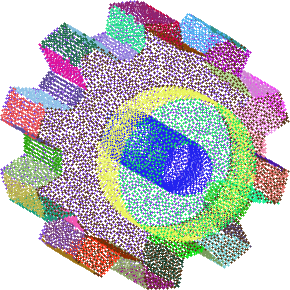}&
      \includegraphics[width=2cm]{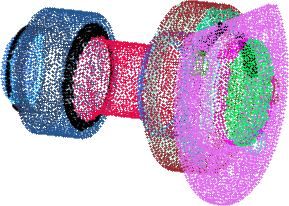}&
      \includegraphics[width=2cm]{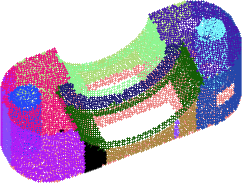}&
      \includegraphics[width=0.85cm]{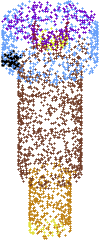}&
      \includegraphics[width=2cm]{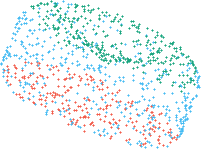}\\ 
      PC $15$ & PC $16$ & PC$17$ & PC $18$ & PC $19$ & PC$20$ & PC $21$ \\
      \hline
      \includegraphics[width=1.2cm]{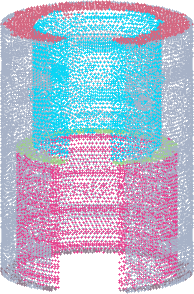}&
      \includegraphics[width=1.8cm]{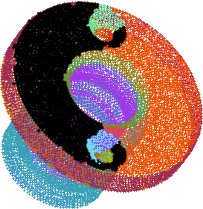}&
      \includegraphics[width=2cm]{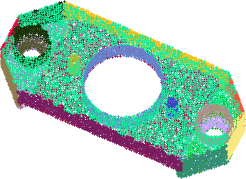}&
      \includegraphics[width=2cm]{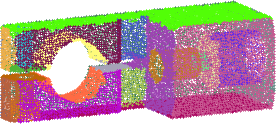}&
      \includegraphics[width=2cm]{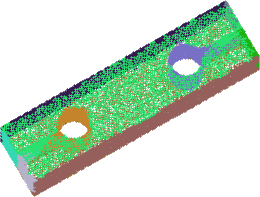}&
      \includegraphics[width=2cm]{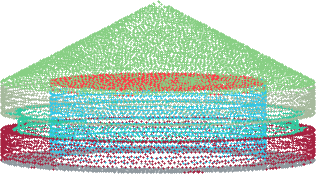}&
      \includegraphics[width=1.6cm]{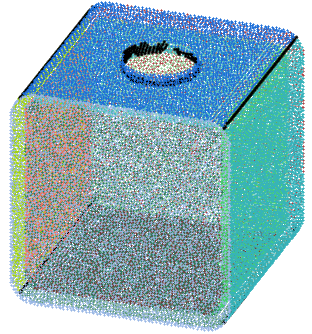}\\ 
      PC $22$ & PC $23$ & PC$24$ & PC $25$ & PC $26$ & PC$27$ & PC $28$ \\
      \hline
        \includegraphics[width=0.9cm]{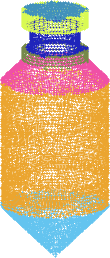}&
      \includegraphics[width=1.8cm]{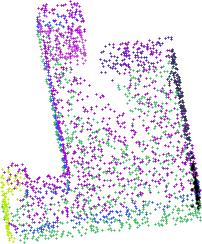}&
      \includegraphics[width=2cm]{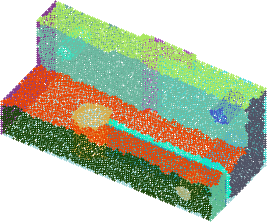}&
      \includegraphics[width=1.5cm]{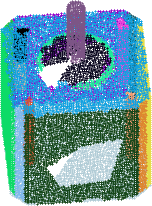}&
      \includegraphics[width=2cm]{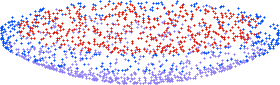}&
      \includegraphics[width=1cm]{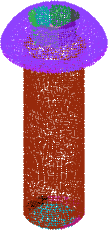}&
      \includegraphics[width=1.8cm]{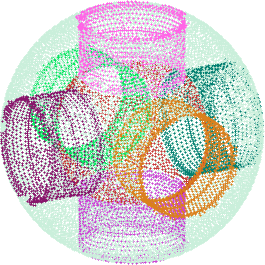}\\ 
      PC $29$ & PC $30$ & PC$31$ & PC $32$ & PC $33$ & PC$34$ & PC $35$ \\
      \hline
    \end{tabular}
    }
    \caption{Segmentations obtained via the PG-based method.}
    \label{fig:res_RANSAC}
\end{figure}

\begin{figure}[h!]
\centering
\resizebox{0.80\textwidth}{!}{
    \centering
    \begin{tabular}{|c|c|c|c|c|c|c|}
    \hline
      \includegraphics[width=2cm]{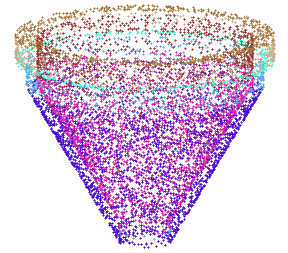}&
      \includegraphics[width=2cm]{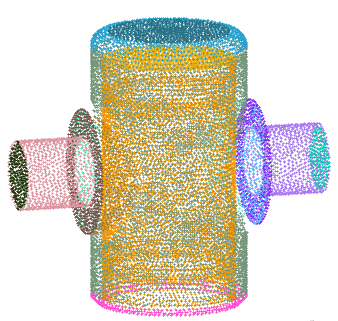}&
      \includegraphics[width=2cm]{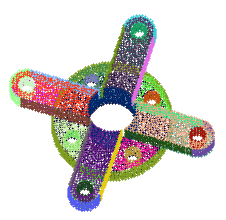}&
      \includegraphics[width=2cm]{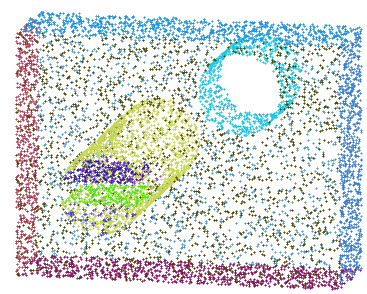}&
      \includegraphics[width=2cm]{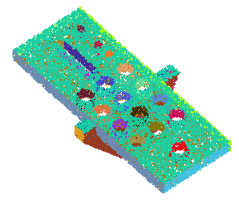}&
      \includegraphics[width=2cm]{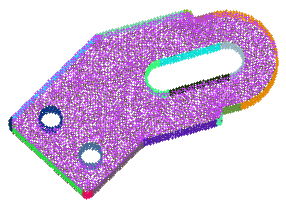}&
      \includegraphics[width=2cm]{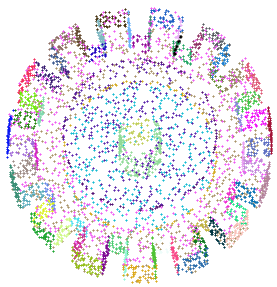}\\ 
      PC $1$ & PC $2$ & PC$3$ & PC $4$ & PC $5$ & PC$6$ & PC $7$ \\
      \hline
       \includegraphics[width=2cm]{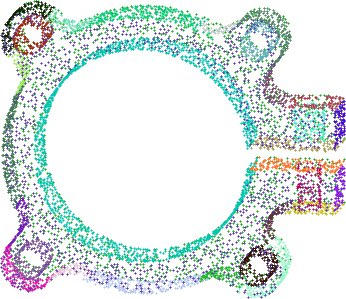}&
      \includegraphics[width=1.5cm]{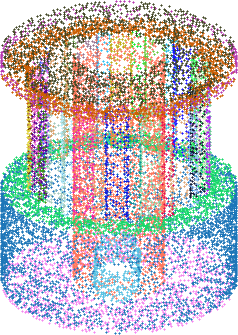}&
      \includegraphics[width=2cm]{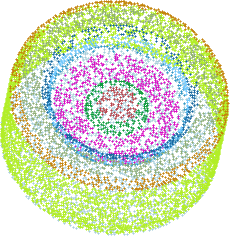}&
      \includegraphics[width=0.8cm]{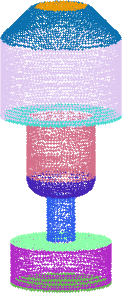}&
      \includegraphics[width=2cm]{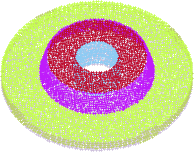}&
      \includegraphics[width=1.8cm]{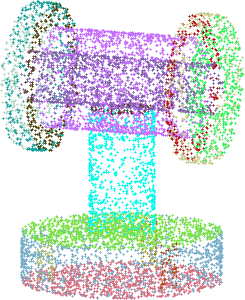}&
      \includegraphics[width=2cm]{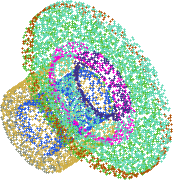}\\ 
      PC $8$ & PC $9$ & PC$10$ & PC $11$ & PC $12$ & PC$13$ & PC $14$ \\
      \hline
      \includegraphics[width=2cm]{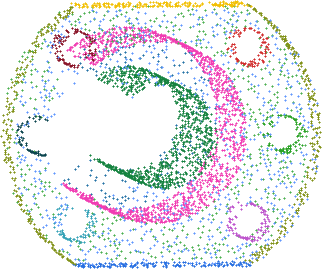}&
      \includegraphics[width=2cm]{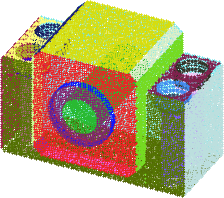}&
      \includegraphics[width=2cm]{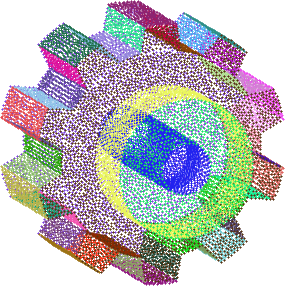}&
      \includegraphics[width=2cm]{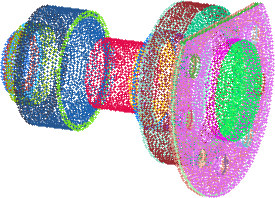}&
      \includegraphics[width=2cm]{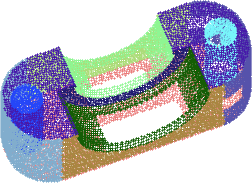}&
      \includegraphics[width=0.85cm]{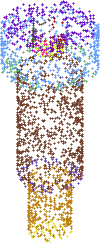}&
      \includegraphics[width=2cm]{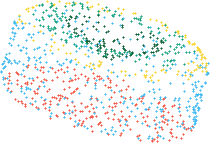}\\ 
      PC $15$ & PC $16$ & PC$17$ & PC $18$ & PC $19$ & PC$20$ & PC $21$ \\
      \hline
      \includegraphics[width=1.2cm]{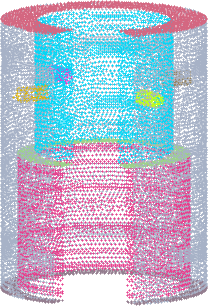}&
      \includegraphics[width=1.8cm]{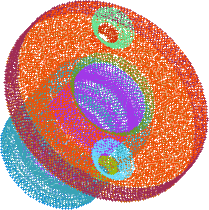}&
      \includegraphics[width=2cm]{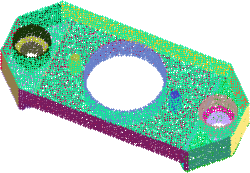}&
      \includegraphics[width=2cm]{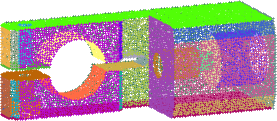}&
      \includegraphics[width=2cm]{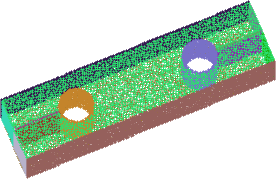}&
      \includegraphics[width=2cm]{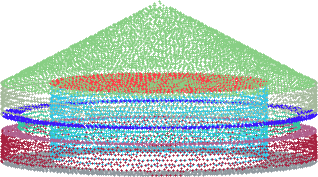}&
      \includegraphics[width=1.6cm]{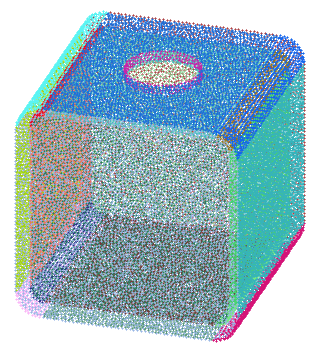}\\ 
      PC $22$ & PC $23$ & PC$24$ & PC $25$ & PC $26$ & PC$27$ & PC $28$ \\
      \hline
        \includegraphics[width=0.9cm]{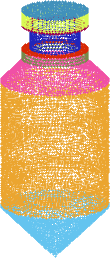}&
      \includegraphics[width=1.8cm]{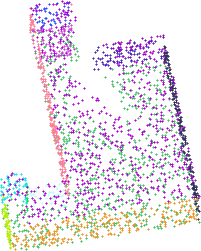}&
      \includegraphics[width=2cm]{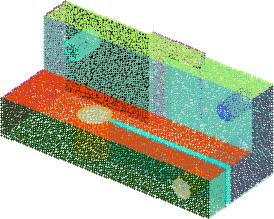}&
      \includegraphics[width=1.5cm]{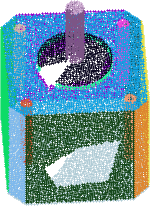}&
      \includegraphics[width=2cm]{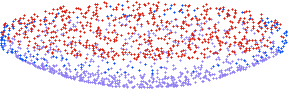}&
      \includegraphics[width=1cm]{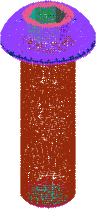}&
      \includegraphics[width=1.8cm]{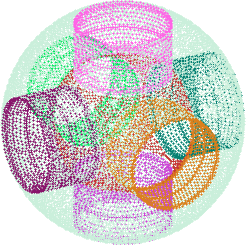}\\ 
      PC $29$ & PC $30$ & PC$31$ & PC $32$ & PC $33$ & PC$34$ & PC $35$ \\
      \hline
    \end{tabular}
    }
    \caption{Segmentations obtained via the Hough-based method.}
    \label{fig:res_HT}
\end{figure}

\end{document}